\documentclass[iop]{emulateapj}
\usepackage{amsmath}
\usepackage{natbib}
\usepackage{graphicx}
\usepackage{epsf}
\usepackage{subfigure}
\usepackage{color}
\usepackage{threeparttable}
\usepackage{comment}
\usepackage{epsfig}
\usepackage{xspace}
\usepackage{enumitem}
\usepackage{hyperref}
\usepackage{ulem}
\usepackage[usenames,dvipsnames,svgnames]{xcolor}
\usepackage{latexsym}

\DeclareGraphicsExtensions{.jpg,.pdf,.pdf,.eps,.ps}

\DeclareSymbolFont{extraup}{U}{zavm}{m}{n}
\DeclareMathSymbol{\varheart}{\mathalpha}{extraup}{86}
\DeclareMathSymbol{\vardiamond}{\mathalpha}{extraup}{87}

\newcommand{\AlensNom}{\ensuremath{1.06 \pm 0.29}}  
\newcommand{\AlensDel}{\ensuremath{0.76 \pm 0.28}}  
\newcommand{\effData}{\ensuremath{28}}              
\newcommand{\delSignificance}{\ensuremath{6.9}}     

\newcommand{\LCDM}{\mbox{$\Lambda$CDM}\xspace}

\def\beq{\begin{equation}}
\def\eeq{\end{equation}}
\def\beqn{\begin{eqnarray}}
\def\eeqn{\end{eqnarray}}
\def\bl{\pmb{\ell}}

\def\ublu{\bl'}
\newcommand{\nhat}{\ensuremath{\mathbf{\hat{n}}}}

\newcommand{\sptpol}{SPTpol\xspace}

\newcommand{\muksq}{\ensuremath{\mu{\rm K}^2}\xspace}
\newcommand{\sqdeg}{\ensuremath{\mathrm{deg}^2}\xspace}

\def\mb{\mathbf}
\def\bhat{\ensuremath{\hat{B}^{\mathrm{lens}}}}
\newcommand\braw{B^{\rm meas}}

\newcommand{\Alens}{\ensuremath{A_{\rm lens}}\xspace}
\newcommand{\Ares}{\ensuremath{A^{\rm res}_{\rm lens}}\xspace}
\newcommand{\Dclbb}{\ensuremath{\Delta C_{\ell}^{BB}}\xspace}

\newcommand{\simleq}{{\raise.0ex\hbox{$\mathchar"013C$}\mkern-14mu \lower1.2ex\hbox{$\mathchar"0218$}}}
\newcommand{\simgeq}{{\raise.0ex\hbox{$\mathchar"013E$}\mkern-14mu \lower1.2ex\hbox{$\mathchar"0218$}}}

\definecolor{orange}{HTML}{BF482F}

\newcommand{\reftext}[1]{\textcolor{Black}{{#1}}}

\newcommand{\refsec}[1]{Section~\ref{sec:#1}}
\newcommand{\refeq}[1]{Eq.~(\ref{eqn:#1})}

\newcommand{\reffig}[1]{Figure~\ref{fig:#1}}

\newcommand{\reftab}[1]{Table~\ref{tab:#1}}

\begin{document}

\title{\vspace{-0.25cm}CMB Polarization B-mode Delensing with SPTpol and Herschel}

\def\KICPChicago{1}
\def\AAUChicago{2}
\def\KIPAC{3}
\def\Stanford{4}
\def\Berkeley{5}
\def\NIST{6}
\def\ArgonneHEP{7}
\def\FNAL{8}
\def\Caltech{9}
\def\JPL{10}
\def\PhysicsUChicago{11}
\def\EFIChicago{12}
\def\UKZN{13}
\def\SLAC{14}
\def\McGill{15}
\def\CIFAR{16}
\def\ColoradoAPS{17}
\def\HarveyMudd{18}
\def\ColoradoPhys{19}
\def\illast{20}
\def\illphy{21}
\def\UChicago{22}
\def\Davis{23}
\def\LBNL{24}
\def\Michigan{25}
\def\Dunlap{26}
\def\ArgonneMSD{27}
\def\Minnesota{28}
\def\Melbourne{29}
\def\CaseWestern{30}
\def\ArtInstChicago{31}
\def\ThreeSpeedLogic{32}
\def\CfA{33}
\def\UToronto{34}
\def\Rochester{35}
\def\aemail{$\dagger$}
\def\ksemail{$\ddagger$}
\def\kwemail{$\sharp$}

\shortauthors{A.~Manzotti, K.~T.~Story, W.~L.~K.~Wu, et al.}
\author{
  A.~Manzotti\altaffilmark{\KICPChicago,\AAUChicago,\aemail},
  K.~T.~Story\altaffilmark{\KIPAC,\Stanford,\ksemail},
  W.~L.~K.~Wu\altaffilmark{\Berkeley,\kwemail},
  J.~E.~Austermann\altaffilmark{\NIST},
  J.~A.~Beall\altaffilmark{\NIST},
  A.~N.~Bender\altaffilmark{\KICPChicago,\ArgonneHEP},
  B.~A.~Benson\altaffilmark{\KICPChicago,\AAUChicago,\FNAL},
  L.~E.~Bleem\altaffilmark{\KICPChicago,\ArgonneHEP},
  J.~J.~Bock\altaffilmark{\Caltech,\JPL},
  J.~E.~Carlstrom\altaffilmark{\KICPChicago,\AAUChicago,\ArgonneHEP,\PhysicsUChicago,\EFIChicago},
  C.~L.~Chang\altaffilmark{\KICPChicago,\AAUChicago,\ArgonneHEP},
  H.~C.~Chiang\altaffilmark{\UKZN},
  H-M.~Cho\altaffilmark{\SLAC},
  R.~Citron\altaffilmark{\KICPChicago},
  A.~Conley\altaffilmark{\NIST},
  T.~M.~Crawford\altaffilmark{\KICPChicago,\AAUChicago},
  A.~T.~Crites\altaffilmark{\KICPChicago,\AAUChicago,\Caltech},
  T.~de~Haan\altaffilmark{\Berkeley},
  M.~A.~Dobbs\altaffilmark{\McGill,\CIFAR},
  S.~Dodelson\altaffilmark{\KICPChicago,\AAUChicago,\FNAL},
  W.~Everett\altaffilmark{\ColoradoAPS},
  J.~Gallicchio\altaffilmark{\KICPChicago,\HarveyMudd},
  E.~M.~George\altaffilmark{\Berkeley},
  A.~Gilbert\altaffilmark{\McGill},
  N.~W.~Halverson\altaffilmark{\ColoradoAPS,\ColoradoPhys},
  N.~Harrington\altaffilmark{\Berkeley},
  J.~W.~Henning\altaffilmark{\KICPChicago,\AAUChicago},
  G.~C.~Hilton\altaffilmark{\NIST},
  G.~P.~Holder\altaffilmark{\CIFAR,\illast,\illphy},
  W.~L.~Holzapfel\altaffilmark{\Berkeley},
  S.~Hoover\altaffilmark{\KICPChicago,\PhysicsUChicago},
  Z.~Hou\altaffilmark{\KICPChicago},
  J.~D.~Hrubes\altaffilmark{\UChicago},
  N.~Huang\altaffilmark{\Berkeley},
  J.~Hubmayr\altaffilmark{\NIST},
  K.~D.~Irwin\altaffilmark{\Stanford,\SLAC},
  R.~Keisler\altaffilmark{\Stanford,\KIPAC},
  L.~Knox\altaffilmark{\Davis},
  A.~T.~Lee\altaffilmark{\Berkeley,\LBNL},
  E.~M.~Leitch\altaffilmark{\KICPChicago,\AAUChicago},
  D.~Li\altaffilmark{\NIST,\SLAC},
  J.~J.~McMahon\altaffilmark{\Michigan},
  S.~S.~Meyer\altaffilmark{\KICPChicago,\AAUChicago,\PhysicsUChicago,\EFIChicago},
  L.~M.~Mocanu\altaffilmark{\KICPChicago,\AAUChicago},
  T.~Natoli\altaffilmark{\Dunlap},
  J.~P.~Nibarger\altaffilmark{\NIST},
  V.~Novosad\altaffilmark{\ArgonneMSD},
  S.~Padin\altaffilmark{\KICPChicago,\AAUChicago,\Caltech},
  C.~Pryke\altaffilmark{\Minnesota},
  C.~L.~Reichardt\altaffilmark{\Melbourne},
  J.~E.~Ruhl\altaffilmark{\CaseWestern},
  B.~R.~Saliwanchik\altaffilmark{\UKZN},
  J.T.~Sayre\altaffilmark{\ColoradoAPS},
  K.~K.~Schaffer\altaffilmark{\KICPChicago,\EFIChicago,\ArtInstChicago},
  G.~Smecher\altaffilmark{\McGill,\ThreeSpeedLogic},
  A.~A.~Stark\altaffilmark{\CfA},
  K.~Vanderlinde\altaffilmark{\Dunlap,\UToronto},
  J.~D.~Vieira\altaffilmark{\illast,\illphy},
  M.~P.~Viero\altaffilmark{\KIPAC},
  G.~Wang\altaffilmark{\ArgonneHEP},
  N.~Whitehorn\altaffilmark{\Berkeley},
  V.~Yefremenko\altaffilmark{\ArgonneHEP},
  and
  M.~Zemcov\altaffilmark{\JPL,\Rochester}
}

\altaffiltext{\KICPChicago}{Kavli Institute for Cosmological Physics, University of Chicago, 5640 South Ellis Avenue, Chicago, IL 60637, USA}
\altaffiltext{\AAUChicago}{Department of Astronomy and Astrophysics, University of Chicago, 5640 South Ellis Avenue, Chicago, IL 60637, USA}
\altaffiltext{\KIPAC}{Kavli Institute for Particle Astrophysics and Cosmology, Stanford University, 452 Lomita Mall, Stanford, CA 94305}
\altaffiltext{\Stanford}{Dept. of Physics, Stanford University, 382 Via Pueblo Mall, Stanford, CA 94305}
\altaffiltext{\Berkeley}{Department of Physics, University of California, Berkeley, CA 94720, USA}
\altaffiltext{\NIST}{NIST Quantum Devices Group, 325 Broadway Mailcode 817.03, Boulder, CO 80305, USA}
\altaffiltext{\ArgonneHEP}{High Energy Physics Division, Argonne National Laboratory, 9700 S. Cass Avenue, Argonne, IL 60439, USA}
\altaffiltext{\FNAL}{Fermi National Accelerator Laboratory, MS209, P.O. Box 500, Batavia, IL 60510}
\altaffiltext{\Caltech}{California Institute of Technology, MS 367-17, 1200 E. California Blvd., Pasadena, CA 91125, USA}
\altaffiltext{\JPL}{Jet Propulsion Laboratory, Pasadena, California , CA 91109, USA}
\altaffiltext{\PhysicsUChicago}{Department of Physics, University of Chicago, 5640 South Ellis Avenue, Chicago, IL 60637, USA}
\altaffiltext{\EFIChicago}{Enrico Fermi Institute, University of Chicago, 5640 South Ellis Avenue, Chicago, IL 60637, USA}
\altaffiltext{\UKZN}{School of Mathematics, Statistics \& Computer Science, University of KwaZulu-Natal, Durban, South Africa}
\altaffiltext{\SLAC}{SLAC National Accelerator Laboratory, 2575 Sand Hill Road, Menlo Park, CA 94025}
\altaffiltext{\McGill}{Department of Physics, McGill University, 3600 Rue University, Montreal, Quebec H3A 2T8, Canada}
\altaffiltext{\CIFAR}{Canadian Institute for Advanced Research, CIFAR Program in Cosmology and Gravity, Toronto, ON, M5G 1Z8, Canada}
\altaffiltext{\ColoradoAPS}{Department of Astrophysical and Planetary Sciences, University of Colorado, Boulder, CO 80309, USA}
\altaffiltext{\HarveyMudd}{Harvey Mudd College, 301 Platt Blvd., Claremont, CA 91711}
\altaffiltext{\ColoradoPhys}{Department of Physics, University of Colorado, Boulder, CO 80309, USA}
\altaffiltext{\illast}{Astronomy Department, University of Illinois at Urbana-Champaign, 1002 W. Green Street, Urbana, IL 61801, USA}
\altaffiltext{\illphy}{Department of Physics, University of Illinois Urbana-Champaign, 1110 W. Green Street, Urbana, IL 61801, USA}
\altaffiltext{\UChicago}{University of Chicago, 5640 South Ellis Avenue, Chicago, IL, USA 60637}
\altaffiltext{\Davis}{Department of Physics, University of California, One Shields Avenue, Davis, CA 95616, USA}
\altaffiltext{\LBNL}{Physics Division, Lawrence Berkeley National Laboratory, Berkeley, CA 94720, USA}
\altaffiltext{\Michigan}{Department of Physics, University of Michigan, 450 Church Street, Ann  Arbor, MI 48109, USA}
\altaffiltext{\Dunlap}{Dunlap Institute for Astronomy \& Astrophysics, University of Toronto, 50 St George St, Toronto, ON, M5S 3H4, Canada}
\altaffiltext{\ArgonneMSD}{Materials Sciences Division, Argonne National Laboratory, 9700 S. Cass Avenue, Argonne, IL 60439, USA}
\altaffiltext{\Minnesota}{School of Physics and Astronomy, University of Minnesota, 116 Church Street S.E. Minneapolis, MN 55455, USA}
\altaffiltext{\Melbourne}{School of Physics, University of Melbourne, Parkville, VIC 3010, Australia}
\altaffiltext{\CaseWestern}{Physics Department, Center for Education and Research in Cosmology and Astrophysics, Case Western Reserve University, Cleveland, OH 44106, USA}
\altaffiltext{\ArtInstChicago}{Liberal Arts Department, School of the Art Institute of Chicago, 112 S Michigan Ave, Chicago, IL 60603, USA}
\altaffiltext{\ThreeSpeedLogic}{Three-Speed Logic, Inc., Vancouver, B.C., V6A 2J8, Canada}
\altaffiltext{\CfA}{Harvard-Smithsonian Center for Astrophysics, 60 Garden Street, Cambridge, MA 02138, USA}
\altaffiltext{\UToronto}{Department of Astronomy \& Astrophysics, University of Toronto, 50 St George St, Toronto, ON, M5S 3H4, Canada}
\altaffiltext{\Rochester}{Center for Detectors, School of Physics and Astronomy, Rochester Institute of Technology, 1 Lomb Memorial Dr., Rochester NY 14623, USA. }

\email{$\dagger$ manzotti@uchicago.edu}
\email{$\ddagger$ kstory@stanford.edu}
\email{$\sharp$ wlwu@berkeley.edu \\}

\begin{abstract}
We present a demonstration of delensing the observed cosmic microwave background (CMB) B-mode polarization anisotropy.
This process of reducing the gravitational-lensing generated B-mode component will become increasingly important for improving searches for the B modes produced by primordial gravitational waves.
In this work, we delens B-mode maps constructed from multi-frequency \sptpol observations of a 90 deg$^2$ patch of sky by subtracting a B-mode template constructed from two inputs: \sptpol E-mode maps and a lensing potential map estimated from the \textit{Herschel} $500\,\micron$ map of the CIB.
We find that our delensing procedure reduces the measured B-mode power spectrum by \effData\% in the multipole range $300 < \ell < 2300$;
this is shown to be consistent with expectations from simulations and to be robust against systematics.
The null hypothesis of no delensing is rejected at $\delSignificance\sigma$.
Furthermore, we build and use a suite of realistic simulations to study the general properties of the delensing process and find that the delensing efficiency achieved in this work is limited primarily by the noise in the lensing potential map.
We demonstrate the importance of including realistic experimental non-idealities in the delensing forecasts used to inform instrument and survey-strategy planning of upcoming lower-noise experiments, such as CMB-S4.
\end{abstract}

\maketitle

\section{Introduction}

In the last two decades, measurements of cosmic microwave background (CMB) anisotropy have played a critical role in establishing \LCDM as the standard model of cosmology (see \citealt{hu02b} for a review).
Increasingly precise measurements of the CMB allow us to test \LCDM with exquisite precision and probe extensions to this model.
In particular, the CMB provides a unique window into the physics of the very early universe.

In the standard cosmological paradigm, the universe underwent a period of near-exponential expansion in its early phase;
this period is called ``cosmic inflation.''
Inflation generically predicts a stochastic background of gravitational waves \citep[see e.g.][for a review]{kamionkowski15}.
These primordial gravitational waves (PGW) imprint a unique signature on the polarized anisotropies of the CMB.
CMB polarization fields can be decomposed into even-parity (divergence) and odd-parity (curl) components, referred to as ``E'' and ``B'' modes, respectively by analogy to the otherwise unrelated properties of electric and magnetic vector fields.
In \LCDM, the PGW background is the only source of B-mode polarization at the epoch of recombination.
The amplitude of this primordial B-mode component is parametrized by the ratio of the amplitudes of the primordial tensor and scalar spectra, $r$, and, in most of the inflationary models, it is directly related to the energy scale of inflation.
Measuring $r$ from CMB B modes provides the cleanest known observational window onto the PGW background.
While the PGW background has not been detected, a variety of arguments predict $r \gtrsim 10^{-3}$, which should be observable in the near future \citep{kamionkowski15}.
Therefore, measuring $r$ is a major objective of current and future CMB experiments.

However, the observed B modes are not solely due to PGWs.
As CMB photons travel to Earth from the last-scattering surface, their paths are deflected by gravitational interactions with large-scale structure (LSS) in the universe, a process known as ``gravitational lensing''.
Lensing shears the CMB polarization pattern, producing ``lensing B modes'' from CMB E modes \citep{zaldarriaga98}.
Lensing B modes were first detected with \sptpol in cross-correlation with LSS \citep{hanson13} and several subsequent detections have been made \citep{polarbear2014c,van-engelen:2015,planck2015XV}.
The (lensing) B-mode auto-spectrum has also now been detected \citep{polarbear2014b,bicep2a,keisler15}.

While these lensing B modes provide valuable information about LSS \citep[e.g.][]{smith08}, they also contaminate searches for the PGW background.
Indeed, current searches for the PGW background are already limited by the contamination from lensing B modes because instrument noise is below the lensing B mode $rms$ fluctuations \citep{bk14}.
A simple way to deal with this is to fit a B-mode power spectrum measurement to a model that includes both components from lensing and primordial tensor modes.
However this method is sub-optimal because the multi-component fit is subject to the sample variance of the lensing B modes.
Instead, the specific realization of lensing B modes on the sky can be characterized and removed in a process called ``delensing,'' thus significantly improving constraints on $r$.
While delensing has been studied on a theoretical level for many years \citep{knox2002, kesden2002, seljak2003, simard:2015,sherwin15,smith:2012},
it has only been performed on CMB data recently \citep{larsen:2016, carron:2017}.

In this paper, we delens a measurement of the CMB B-mode power spectrum using 90 deg$^2$ of \sptpol data.
We build a template of the lensing B modes on the sky from the \sptpol E-mode map and an estimate of the CMB lensing potential constructed from a \textit{Herschel} 500$\mu m$ map of the cosmic infrared background (CIB).
This B-mode template is subtracted from the measured \sptpol B-mode maps, and the resulting maps are used to calculate the delensed power spectrum.

In parallel, we have developed a suite of realistic simulations that incorporate sky components, instrumental and data-processing effects, and instrumental noise.
These simulations allow us to interpret our results and to test their robustness against possible systematic effects in the data or in the assumptions required to construct a B-mode template.
Furthermore, we use these simulations to investigate the factors that affect the efficiency of the delensing procedure.

During the final stage of preparing this paper for publication, we became aware of \cite{carron:2017} in which the authors also delens B-mode polarization maps.
Our work differs from \cite{carron:2017} in the delensing methodology:
whereas \cite{carron:2017} use a displacement field estimated from the internally measured CMB lensing potential to undo the lensing-induced deflection in CMB temperature and polarization maps,
we form a B-mode template from maps of E modes and the CIB-estimated CMB lensing potential then subtract this template from measured B-mode maps.
In addition, our work investigates non-ideal experimental effects in the input maps that degrade the delensing efficiency.

This paper is structured as follows:
\refsec{theory} briefly introduces the theoretical framework, \refsec{data} describes the instrument, observations, and data products used in the analysis, \refsec{processing} explains the analysis process, and \refsec{sim} describes the suite of simulations. The main results are presented in \refsec{res} and systematics, null, and consistency tests are shown in \refsec{sys-null-sign}.
A study of delensing efficiency is presented in \refsec{discussion}.
We conclude in \refsec{conclusion}.

\setcounter{footnote}{0}
\section{Lensing B modes and Our Approach to Delensing}
\label{sec:theory}
On their way from the last scattering surface to us, CMB photons are gravitationally deflected by the large-scale structure of the universe.
This leads to a remapping of the observed CMB fields as:
\beq
X(\hat{\mathbf{n}}) = X_{\mathrm{unlensed}}(\hat{\mathbf{n}}+\mathbf{d});
\eeq
where $X$ can be either the temperature field T or one of the two polarization Stokes parameters Q and U, and $\mathbf{d}$ is the lensing deflection field, which is given by the gradient of the lensing potential $\mb{d}(\nhat) = \nabla \phi(\nhat)$.
The 2d potential $\phi$ is a weighted integral of the 3d gravitational potential along the line of sight and has contributions from a broad range of redshifts that peaks at $z\sim 2$ (see, e.g. Figure 1 of \citealt{planck2013XVIII}).

The CMB polarization fields can be conveniently described in terms of even-parity E modes and odd-parity B modes.
Lensing deflections shear E modes, producing a B-mode lensing component.
In the flat-sky approximation\footnote{
In this paper, we use the flat-sky approximation to relate multipole number $\ell$ to $\mb{u}$, the Fourier conjugate of the Cartesian angle on a small patch of sky, as $\bl = 2\pi\mb{u}$ and $\ell =2\pi|\mb{u}|$.},
this lensing component is given to first order in $\phi$ by:
\beq\label{eqn:blens}
B^{\mathrm{lens}}(\bl) =  \int \frac{d^2 \bl'}{(2 \pi)^2} W(\bl,\bl') E(\ublu) \phi(\bl - \bl')
\eeq
with
\beq
W(\bl,\bl') = \bl' \cdot (\bl-\bl') \sin(2\varphi_{\bl,\bl'}) \,,
\eeq
where $\varphi_{\bl,\bl'}$ is the angle between the two vectors $\bl$ and $\bl'$.

Lensing B modes obscure PGW B modes, thus it is important to model and remove the lensing B-mode component.
We do this by building a template of the lensing B modes in the observed patch from $\bar{E}$, a filtered E-mode map,
and $\hat{\phi}(\bl)$, an estimated map of the lensing potential.
To first order in $\phi$, we can optimally estimate the lensing B modes as:
\beq \label{eqn:blens_estimate}
\bhat(\bl) = \int \frac{d^2 \bl'}{(2 \pi)^2} W(\bl,\bl') \bar{E}(\ublu) \hat{\phi}(\bl - \bl') \,.
\eeq
Both $\bar{E}(\bl)$ and $\hat{\phi}(\bl)$ are filtered:
The E-map filtering described in \refsec{btemplate} is a Wiener filter of the (noisy) E-map, $E^N$.
This filter can be approximated as
\beq \label{eqn:fle}
\bar{E}(\bl) \approx \left(\frac{C^{EE}_{{\ell}}}{C^{EE}_{{\ell}}+N^{EE}_{{\ell}}} \right) E^N(\bl) \,,
\eeq
when the signal and noise are isotropic in 2d Fourier space.
Here $C^{EE}_{\ell}$ and $N^{EE}_{\ell}$ are the theoretical 1d E-mode signal and noise power spectra, respectively.

In this analysis, we estimate $\phi$ from a map of the CIB.
The CIB has contributions out to high redshift and thus traces the CMB lensing potential quite well \citep{Song:2002sg,Holder:2013hqu,planck2013XVIII}.
For now, the CIB provides a higher signal-to-noise estimate of the lensing potential than reconstructing $\phi$ directly from the CMB \citep{sherwin15}, though with improved data, CMB reconstructions will ultimately provide the best maps of the CMB lensing potential.
The estimate of the projected potential is also filtered.
A filtered estimate of the lensing potential is obtained from a CIB map $I^{\rm CIB}(\bl)$ as
\beq \label{eqn:flp}
\hat{\phi}^{\rm CIB}_{\bl} = \left(\frac{C_l^{{\rm CIB}\mbox{-}\phi}} {C_l^{{\rm CIB}\mbox{-}{\rm CIB}}} \right) I^{\rm CIB}(\bl) \,,
\eeq
where $C_l^{{\rm CIB}\mbox{-}\phi}$ is the cross-spectrum of the CIB and $\phi$, and $C_l^{{\rm CIB}\mbox{-}{\rm CIB}}$ is the auto-spectrum of the CIB (see \refsec{cib} for details about the specific filters applied in this work).
The filters from \refeq{fle} and (\ref{eqn:flp}) formally minimize the residual lensing B modes \citep{sherwin15}.

\newcommand\res{del}
We delens our B-mode maps by subtracting this estimate $\bhat(\bl)$ from the measured B-mode maps.
If the measured B-mode maps contained only the lensed noisy E-mode signal, then the residual map, the {\it delensed} B-mode map, would be
\begin{small}
\beqn
\label{eqn:Bres}
B^\mathrm{\res}(\bl) &=&  \braw(\bl) - \bhat(\bl) \\
 &=&  \int \frac{d^2 \bl'}{(2 \pi)^2} W(\bl,\bl') \times \nonumber \left[ E(\ublu) \phi(\bl-\bl') - \bar{E}(\ublu) \hat{\phi}(\bl-\bl') \right]
\eeqn
\end{small}
Using the approximations of \refeq{fle} and \refeq{flp}, the residual lensing B-mode power becomes \citep{sherwin15}:
\beqn \label{eqn:clbb_res}
C_{\ell}^{BB \mathrm{,\res}} &=&  \int \frac{d^2 \bl'}{(2 \pi)^2} W^2(\bl,\bl')  C_{\ell'}^{EE} C^{\phi\phi}_{|\bl-\bl'|} \nonumber \\
  && \times \left [ 1 - \left(  \frac{C_{\ell'}^{EE}}{C_{\ell'}^{EE} + N_{\ell'}^{EE}} \right ) \rho^2_{|\bl-\bl'|}  \right]
\eeqn
where the CIB-$\phi$ correlation coefficient is given by
\beq \label{eqn:rho}
\rho_{\ell} = \frac{C_{\ell}^{{\rm CIB}\mbox{-}\phi}} {\sqrt{C_{\ell}^{{\rm CIB}\mbox{-}{\rm CIB}} C_{\ell}^{\phi\phi}}}.
\eeq
Theoretically, this produces the delensed B-mode map with the minimum possible variance \citep{sherwin15}.
\reffig{CIB} shows $\rho_{\ell}$ for the assumed model detailed in \refsec{cib}.

\section{Data}
\label{sec:data}
\subsection{CMB Data: The \sptpol Instrument and Observations}
\label{sec:instrument}
The South Pole Telescope is a 10 meter diameter telescope located at the Amundsen-Scott South Pole Station in Antarctica \citep{carlstrom11, padin08}.
During the austral summer of 2011-2012, the polarization-sensitive \sptpol receiver was installed on the SPT.
\sptpol contains 1536 transition-edge sensor bolometers; 1176 detectors at 150 GHz and 360 detectors at 95 GHz.
For details of the receiver, see \citet{crites14} (hereafter C15) and references therein.

From April to October 2012 and in April 2013, \sptpol was used to observe a 100 deg$^2$ patch of sky centered at right ascension 23h30m and declination $-$55 degrees.
We refer to this field as the SPTpol ``100d'' field to distinguish it from the 500~\sqdeg\ survey field,
for which observations began in May 2013.
The 100d field spans from 23h to 0h in right ascension and from $-$50 to $-$60 deg in declination.
The data collected from this field has been used to make the first detection of CMB B modes \citep[][hereafter H13]{hanson13},
measure the temperature and E-mode anisotropy power spectrum (C15),
measure the B-mode anisotropy power spectrum \citep[][hereafter K15]{keisler15},
and measure the CMB lensing potential \citep[][hereafter S15]{story14}.
Details of these observations can be found in those references; we summarize the salient parts here.

The field was observed by scanning the telescope back and forth once in azimuth at a fixed elevation, followed by a small step in elevation.
This scan pattern was repeated until the full extent of the field in declination was covered.
The field was observed using a ``lead-trail'' strategy in which half of the extent in right ascension is observed first and then the second half is observed so that during the observation the telescope covers an identical range in azimuth; see C15 or K15 for details.

We refer to a pass of the telescope from one side of the field to the other as a ``scan,'' and to one set of scans covering the complete field as an ``observation.''
This analysis uses the same set of $\sim 6000$ observations as K15, which corresponds to roughly 6000 hours of observation.

\subsection{\textit{Herschel} cosmic infrared background}
\label{sec:cib}
In this work, the CMB lensing potential is estimated from maps of the cosmic infrared background (CIB).
We use CIB maps obtained with the SPIRE instrument \citep{Griffin:2010hp} on board the \textit{Herschel} space observatory \citep{Pilbratt:2010mv}.
These maps are presented in more detail in H13, and have 90 deg$^2$ of overlap with the 100d field (see \refsec{bb_maps}).

\begin{figure}
\begin{center}
\includegraphics[width=0.48\textwidth]{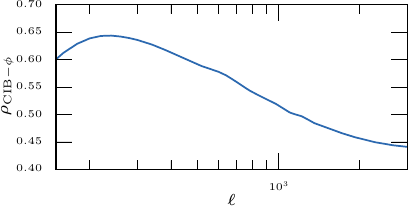}
\caption{Correlation coefficient $\rho_{\ell} = \frac{C_{\ell}^{{\rm CIB}\mbox{-}\phi}} {\sqrt{C_{\ell}^{{\rm CIB}\mbox{-}{\rm CIB}} C_{\ell}^{\phi\phi}}}$ between the lensing potential $\phi$ and the CIB. The model used to compute this is described in \refsec{cib}. Depending on the angular scale, we expect $45\%-65\%$ of the CIB to be correlated with $\phi$. 
\reftext{Almost all angular scales below $\ell<1000$ in $\phi$ (with a relative weight gradually decreasing at high $\ell$) are responsible for generating lensing B-modes in the range considered in this work (see Fig. 2 in \citealt{simard:2015}).}
}
\label{fig:CIB}
\end{center}
\end{figure}

We estimate the CMB lensing potential $\hat{\phi}^{\rm CIB}$ from the \textit{Herschel} $500\,\micron$ map since it has the best overlap with the CMB lensing kernel of the three SPIRE bands \citep{Holder:2013hqu}.
The raw $500\,\micron$ \textit{Herschel} map is apodized and Fourier-transformed, then all the modes $\ell\!\le\!150$ are removed to avoid Galactic dust contamination.
We then apply the Wiener filter of \refeq{flp}.
Following \cite{Addison:2011se}, we model the CIB power as $C_{\ell}^{{\rm CIB}\mbox{-}{\rm CIB}} = 3500 (l/3000)^{-1.25} {\rm Jy^2 / sr}$.
We test that this model provides an accurate fit for the power of the \textit{Herschel} $500\,\micron$ map used in this work.
For the cross-spectrum $C_{\ell}^{{\rm CIB}\mbox{-}\phi}$, we use the single-SED model of~\cite{hall:2010}.
This places the peak of the CIB emissivity at redshift $z_c = 2$ with a broad redshift kernel of width $\sigma_z = 2$.
This model is rescaled to agree with the results of
~\citet{Holder:2013hqu} and \citet{planck2013XVIII} by choosing the corresponding linear bias parameter.
Other multi-frequency CIB models are available \citep[e.g.,][]{Bethermin:2013nza}; however, given the level of noise, we are relatively insensitive to this choice.
With these assumptions, depending on angular scale, $45-65\%$ of the CIB is correlated with the CMB lensing potential, as shown in \reffig{CIB}.

We show in \refsec{sys} that the delensing efficiency is robust against reasonable changes in the assumed ${\rm CIB}\mbox{-}\phi$ correlation.
In \refsec{discussion}, we explore how much the delensing efficiency improves with a higher CIB-$\phi$ correlation.

\section{Data Processing}
\label{sec:processing}
This analysis requires two related data-processing pipelines to process time-ordered telescope data into maps:
the first pipeline produces the B-mode maps that are used to calculate the B-mode power spectrum,
while the second pipeline produces E-mode maps that are then combined with the lensing field $\hat{\phi}^{\rm CIB}$ to form the lensing B-mode template.
This section first describes a common pipeline to produce temperature (T) and polarization maps (Q,U) from bolometer data, followed by separate sections describing the two pipelines producing the B-mode maps and the lensing B-mode template.
We conclude by briefly describing how we compute the final power spectrum from these maps.

\subsection{Bolometer data to Maps}
\label{sec:processing_maps}
This section describes how time-ordered data are processed into T, Q, and U maps up to where the two pipelines diverge.
The data and processing described in this section are identical to that of K15.
The process is summarized here (see C15 and K15 for details).

Time-ordered data from individual bolometers are filtered by removing a first-order polynomial (mean and slope) and a set of low-order Fourier modes from each scan.
Because scans lie at constant declination, this filtering acts as a high-pass filter in the direction of right ascension with an effective cutoff of $\ell_x \sim 100$.
We mask all sources with flux above 50 mJy during filtering to avoid filtering artifacts.
Contamination in the time-ordered data from the frequency of the pulse tube cooler is removed with a Fourier-domain notch filter.

As in K15, we calibrate bolometer timestreams relative to each other using a combination of observations of an internal chopped blackbody source and of the galactic HII region RCW38.
We measure the polarization angle and efficiency of each detector from observations of a far-field chopped thermal source behind a rotating wire grid.
We flag and remove data from individual bolometers on a per-scan and a per-observation basis based on poor noise performance, discontinuities in the data, and unusual response to the calibrator observations.

The filtered, calibrated bolometer data are then combined and binned into T, Q, and U maps according to the process described in K15.
Maps are made using the oblique Lambert azimuthal equal-area projection with 1-arcmin pixels.
The maps are in units of $\mu {\rm K}_{\rm CMB}$, corresponding to the temperature variations in a 2.73 K blackbody that would produce the observed variations in intensity.

The absolute calibration of these maps is obtained by the process described in C15,
in which we calculate the cross-power spectrum between the 150 GHz maps used here and the 150 GHz SPT-SZ maps of the same field,
which were in turn calibrated by comparing the temperature power spectrum reported in \cite{story13} to the \cite{planck2013XVI} temperature power spectrum.
The calibration uncertainty resulting from this process is $1.3\%$ in temperature for both the 150 GHz and 95 GHz maps and is almost $100\%$ correlated between these frequencies.
The absolute polarization calibration is calculated by comparing the $EE$ spectrum measured in K15 to the $EE$ spectrum from the best-fit \LCDM model for the
\textsc{plikHM\_TT\_lowTEB\_lensing} constraint from \cite{planck15-11}.

This processing results in a T, Q, and U map for each observation of the field.
We use the 95 and 150 GHz maps from the K15 analysis, for which the effective white noise levels are
approximately 17 and 9~$\mu$K-arcmin in polarization.

\subsection{B-mode map pipeline}
\label{sec:bb_maps}
 This section describes how we derive B-mode maps from the T, Q, and U maps.
These will be the maps denoted as $\braw$ in \refeq{Bres}.
This pipeline is identical to the approach in K15, except the \textit{Herschel} sky coverage is incorporated into the apodization mask.

The single-observation maps from the previous section are combined into 41 sets of maps with relatively uniform map coverage which we refer to as ``bundles;'' we use the same bundle definitions as K15.
The B-mode power spectrum described in \refsec{clbb} is calculated from the cross-spectrum between these bundles in order to avoid noise bias \citep{polenta05,tristram05}.

The maps are apodized to down-weight high-noise regions near the edges of the maps and restrict the data to the sky covered by the CIB data.
The valid-data region is the intersection of the non-zero regions of the K15 apodization mask and the CIB mask.
This valid-data region is apodized using a Gaussian smoothing kernel with $\sigma = 10'$ (the same process used for K15).
This apodization mask has an effective area of $\sim 90$ deg$^2$ -- 11 \% smaller than the K15 analysis.

We perform two map-space cleaning steps.
First, to reduce contamination from emissive point sources, we interpolate over all sources with unpolarized flux $S_{150}>50$ mJy.
Second, the maps are filtered to reduce the effects of scan-synchronous signals exactly as described in K15.

Taking the Fourier transform of each bundle, a harmonic-space $B_{\bl}$ map is constructed from the corresponding $\{Q,U\}$ map pair using the $\chi_B$ method described in \cite{smith07b}.
These $B_{\bl}$ map bundles form the input to the cross-spectrum analysis described below in \refsec{clbb}.

\subsection{B-template map pipeline}
\label{sec:btemplate}
This section describes how we use the maps from \refsec{processing_maps} to produce a lensing B-mode template.
This processing has many similarities to the pipeline used in S15; we refer the reader to S15 for additional details.

The maps from \ref{sec:processing_maps} are first coadded into a single T, Q, and U map set (i.e., not split into bundles).
The T, Q, and U maps are masked with a binary mask which is the intersection of the valid-data region (described in \refsec{bb_maps}) with the source mask from S15 (described in Section 2 of S15).
Next we calculate the harmonic space filtered $E_{\bl}$ map using the filtering process described in Section 3.1 of S15; see that reference for details.
This process is essentially a ``matched filter'' designed to maximize the expected signal-to-noise of CMB modes in 2d Fourier space,
and it approaches the simple form of \refeq{fle} in ideal conditions.
This process produces a filtered harmonic-space map $\bar{E}_{\bl}$.
Note, we use only the 150 GHz E-mode data to form the B template.

\begin{figure*}
\begin{center}
\includegraphics[width=1\textwidth]{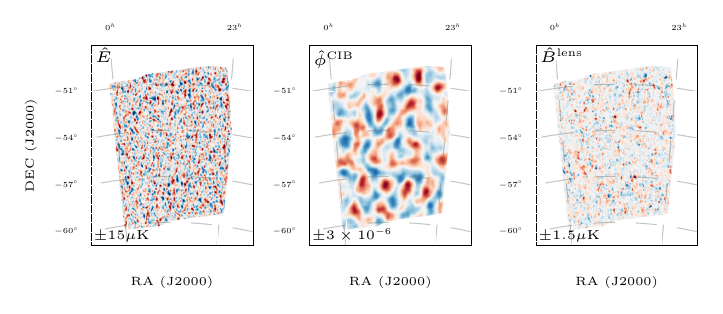}
\caption{Maps used for delensing the data.
The filtered E-mode map $\bar{E}^{150}$ (left) is combined with a tracer of the CMB lensing potential obtained from filtering the CIB $\hat \phi^{\rm{CIB}}$ (center) in Fourier space to obtain a template of the lensing B modes $\hat B^{\rm{lens}}$ (right). This template is then subtracted from the B-mode data.}
\label{fig:data-maps}
\end{center}
\end{figure*}

The lensing B-mode template \bhat{} is formed from $\bar{E}_{\bl}$ and $\hat{\phi}^{\rm CIB}_{\bl}$ using \refeq{blens_estimate};
these three maps are shown in \reffig{data-maps}.

It is important to note that since the input fields $\bar{E}_{\bl}$ and $\hat{\phi}^{\rm CIB}$ are filtered, \bhat{} is a \textit{filtered} estimate of the lensing B modes in this patch of sky.
The goal of delensing is to minimize the variance of the residual lensing B modes after delensing.
The filtered B template accomplishes this goal at the cost of not recovering all of the lensing power.
In other words, \bhat{} does not have unit-response to lensing B modes.
This can be seen in \reffig{bhat_auto}, where none of the simulated B template auto-spectra recover the full input lensing power.
If this filtering were not applied, \bhat{} would be noise-dominated and would significantly \textit{increase} the variance of the delensed power spectrum relative to the nominal spectrum\footnote{\textit{Nominal} here and throughout the paper refers to the normal power spectrum, without any delensing.}, in our case by almost an order of magnitude.

\subsection{Power Spectrum Analysis}
\label{sec:clbb}
With B-mode maps (from \refsec{bb_maps}) and the B-mode template \bhat{} (from \refsec{btemplate}) in hand,
we proceed to calculate the nominal and delensed B-mode power spectra $C_{\ell}^{BB}$.
We refer to the power in each bin in $\ell$ as a ``bandpower.''

The power spectrum is calculated using the K15 analysis pipeline, which implements a ``pseudo-Cl'' cross-spectrum method \citep{hivon02}.
The harmonic-space B-map bundles are passed through a number of steps: convolutional cleaning, calculating the cross-spectrum between all bundle pairs, applying a 2d Wiener filter, averaging and binning into a 1d spectrum, subtracting an additive bias, and applying a multiplicative bias correction.
This process produces three sets of bandpowers: 95 GHz and 150 GHz auto-spectra, and a $95 \times 150$ GHz cross-spectrum.
We also use the inverse-variance-weighted combination of these three sets of bandpowers, referred to as the ``spectrally-combined'' bandpowers.
Because we use the same pipeline as K15, we refer the reader to that paper for details.
The delensing process differs only in that the B-mode template \bhat{} is subtracted from each bundle of B maps before calculating the cross-spectra.

Modes in the B-map bundles have been suppressed by the filter transfer function, while the modes in \bhat{} are not equally suppressed.
Subtracting \bhat{} directly from the B-map bundles would over-subtract modes reduced by filtering and add power to the final delensed power spectrum.
To account for this, we multiply \bhat{} by the 2d filter transfer function estimated from simulations, then subtract this ``biased'' version of \bhat{} from the B-map bundles.
The multiplicative bias correction then accounts for these suppressed modes.

In order to quantify the power removed by delensing, the B-mode spectra (both nominal and delensed) in our main results presented in \refsec{res} are fit to the expected foreground-free theoretical spectrum scaled by a single amplitude \Alens following K15.
As in K15, this fit is calculated from the two auto-spectra and the cross-spectrum of different frequencies (5 bandpowers each, for a total of 15 data points) using the variance of each bandpower from noisy simulations.
The amplitude of the nominal bandpowers is denoted \Alens; the amplitude of the residual delensed bandpowers is denoted \Ares.

We define the ``delensing efficiency'' from this amplitude as the percent of lensing power removed with the delensing procedure:
\beq
\label{eqn:efficiency}
  \alpha = \frac{\Alens - \Ares}{\Alens} \,.
\eeq
The efficiency will approach one for perfect delensing and zero for no delensing.

Finally, it is useful to consider the difference between the nominal and delensed bandpowers.
This ``spectrum difference'' is defined as
\beq
\label{eqn:diff}
  \Delta C_{\ell}^{BB} \equiv C_{\ell}^{BB}-C_{\ell}^{BB \mathrm{,\res}}
\eeq
and is the amount of power \textit{removed} by delensing.

\section{Simulations}
\label{sec:sim}

This analysis and its interpretation depends critically on an accurate and realistic suite of simulations.
Simulated skies are formed from lensed CMB and foreground emission components.
These skies are then passed through a ``mock-observing'' pipeline to simulate the effects of \sptpol observations and data processing.
This gives us an accurate and realistic suite of simulations.

In \refsec{sys-null-sign} we use simulations to quantify the significance of our results and to test their robustness against possible systematics in the data.
These simulations are used in \refsec{discussion} to separate out different factors affecting delensing efficiency and to understand where improvements in efficiency can be expected in the future.

In this section, we first describe how the simulated CMB and CIB skies are generated.
We then discuss several different simulated B-mode templates that will be used to understand the delensing efficiency.

\begin{figure*}
\begin{center}
\includegraphics[width=0.95\textwidth]{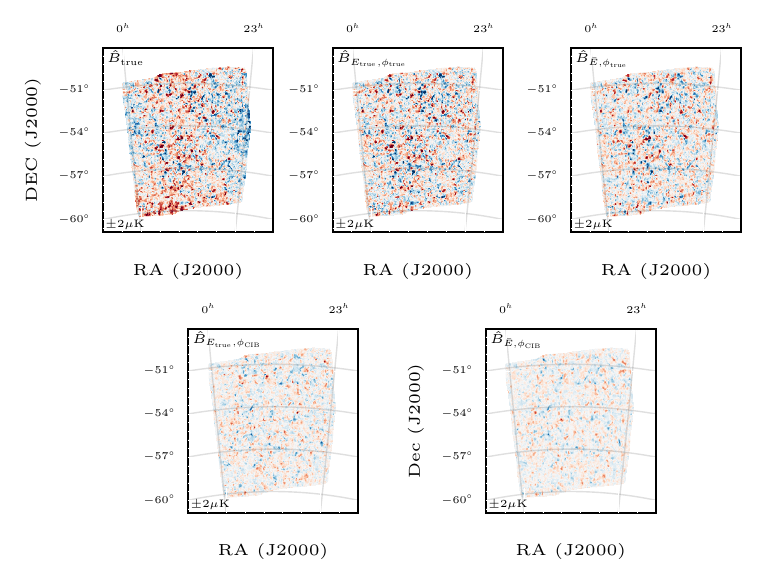}
\caption{Simulated B-mode templates for visualizing delensing efficiency in map space.
  This figure shows how increasing levels of noise in our simulations degrade our ability to recover lensing B modes.
  The different cases are labeled in the top-left corner and described in \refsec{sim_template}.
  These are real space maps, even though our analysis is performed in Fourier space.
  \textbf{Top:} True B-mode map (left), the template constructed from true E and true $\phi$ (center), and the template constructed from filtered E and true $\phi$ (right).
  \textbf{Bottom:} B-mode templates built from noisy $\phi$'s estimated from CIB, $\phi_{\rm CIB}$, in combination with true E (left) and filtered E (right).
  The bottom-right ``realistic template'' ($\hat{B}^{150}_{\bar{E},\phi_{\rm CIB}}$) should accurately simulate the actual B-mode template estimated from data.
}
\label{fig:templ-maps}
\end{center}
\end{figure*}

\subsection{Pipeline}
\label{sec:sim_pipeline}
We generate realizations of un-lensed CMB anisotropies (T,Q,U) and the lensing potential from the fiducial cosmological model.
Our fiducial cosmology is the \LCDM model that best fits the 2015 \textsc{plikHM\_TT\_lowTEB\_lensing} dataset \citep{planck15-11}.
The CMB skies are then lensed using realizations of the lensing potential using \textsc{Lenspix} \citep{lewis05}.
At this step, the lensed CMB skies and lensing potential maps are projected directly into the format of the 100d \sptpol map.
The resulting ``truth'' maps are referred to as $E_{\rm true}$, $B_{\rm true}$, and $\phi_{\rm true}$.

We next add a Gaussian realization of our foreground model to each simulated CMB sky.
The components of this model are taken from measured values where known, and upper limits otherwise.
In the temperature skies we add several components: the thermal Sunyaev-Zel'dovich (tSZ), the clustered and unclustered components of the CIB (dusty sources), and radio point sources (AGN).
The tSZ component is modeled by a tSZ power spectrum template taken from \cite{shaw10} rescaled by $A_{\rm tSZ}$.
We use $A_{\rm tSZ} = 4~\muksq$ for 150GHz and $A_{\rm tSZ} = 12~\muksq$ for 95GHz \citep{george14}.
The other three sources are modeled by power-laws in angular multipole $\ell$ space with the form:
\begin{equation*}
D_{\ell, \mathrm{source}}^{i}=  A^{i}_{\rm source}\left(\frac{\ell}{3000}\right)^{p} \,,
\end{equation*}
where $i\in\{150 \rm{GHz} ,95 \rm{GHz} \}$ and $D_{\ell} = \frac{\ell(\ell+1)}{(2\pi)}C_{\ell}$.
For the clustered CIB term, we use $p=0.8$ and  $A_{\rm{CIB}}^{95 \rm{GHz}} = 0.56~ \muksq$ and $A_{\rm{CIB}}^{150 \rm{GHz}} = 3.46~ \muksq$  \citep{george14}.
We neglect the correlation between these CIB components at 150 and 95 GHz and the simulated CIB map at $500 \mu m$.
Power from unclustered point sources is by definition flat in $C_{\ell}$, so $p=2$ for unclustered sources.
With the adopted threshold for point source masking of 50~mJy at 150GHz, the residual dusty sources power spectrum amplitudes are
$A_{\rm{dusty}}^{150 \rm{GHz}} = 9~\muksq$ and $A_{\rm{dusty}}^{95 \rm{GHz}} = 1.5~\muksq$ \citep{george14};
and the residual radio sources amplitudes are
$A_{\rm{radio}}^{150 \rm{GHz}} = 10~ \muksq$ and $A_{\rm{radio}}^{95 \rm{GHz}} = 50~\muksq$ \citep{mocanu13, george14}.
We need the foreground component in the temperature maps to properly account for T/E $\rightarrow$ B leakage that is removed through the convolutional cleaning step of the power spectrum analysis of the B-mode map (\refsec{clbb}).

We model the polarized foregrounds in this work with two components: galactic and extragalactic.
Following the arguments from K15, we model the contribution from polarized galactic dust as
\begin{equation*}
  D_{\ell,\mathrm{dust}}^{X,i} = A_{\rm dust}^{X,i} \left(\frac{\ell}{80}\right)^{-0.42}
\end{equation*}
for $X\in\{E,B\}$ and $i \in\{95,150\}$ GHz.
As in K15, we use the values $A_{\rm dust}^{B, 150 \rm GHz} = 0.0118\muksq$ and $A_{\rm dust}^{B, 95 \rm GHz} = 0.00169\muksq$,
and the value for the EE dust spectra is twice that of the BB spectra in both frequencies \citep{2016A&A...586A.133P}.
We model the extragalactic polarized sources by assigning polarization fractions to unclustered sources:
We set the polarization fraction to 2\% for dusty sources (based on \citealt{seiffert07}) and 3.7\% for radio sources (based on an investigation of the polarization properties of bright AGN in the SPTpol 500d survey field).

The process above produces a set of simulated CMB+foreground sky signal maps.
In order to simulate the effects of the \sptpol instrument and data processing, these simulated skies are passed through a mock observation and data processing pipeline.
First, the sky maps are convolved with the azimuthally symmetric beam function from K15.
Next, \sptpol pointing information from each observation is used to create bolometer time-ordered data for each beam-convolved simulation map.
These data are then processed into simulated maps on a flat-sky projection using the data pipeline as described in \refsec{processing_maps}.

As a final step, noise realizations are added to the simulations.
These noise realizations are estimated directly from the data as described in K15.
We make noise realizations for each simulated map-bundle of a simulated sky by
splitting the input data maps from that bundle into two random halves, coadding each half, then subtracting these coadds.
These are noise realizations for each map-bundle in the B-mode map pipeline.
For the B-template map pipeline, we coadd the noise realizations of the map-bundles for each simulated sky
and add the coadded noise realizations to the simulated Q/U maps.

The CIB skies are generated in the same way as in H13.
For each CMB lensing potential realization $\phi(\nhat)_i$, we generate a Gaussian map that has the correlation with CMB lensing corresponding to $C_{\ell}^{{\rm CIB}\mbox{-}\phi}$ described in \refsec{cib}.
We then add Gaussian noise, uncorrelated with the CMB lensing field, such that the total power for the CIB is equal to $C_{\ell}^{{\rm CIB}\mbox{-}{\rm CIB}}$.

For this work we generated 100 CIB skies and lensed CMB simulations.
We also use 100 foreground-free unlensed simulations to characterize the additive bias at the power spectrum analysis step.
These simulated maps are then passed through both the B-mode map pipeline from \refsec{processing_maps} and the B-template map pipeline from \refsec{bb_maps},
allowing us to calculate B-mode bandpowers with and without delensing.

\subsection{Simulated B-mode templates}
\label{sec:sim_template}

\begin{figure}
\begin{center}
\includegraphics[width=0.45\textwidth]{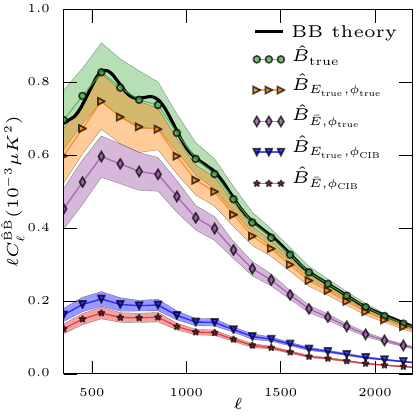}
\caption{The distribution of (filtered) B-mode template auto-spectra from simulations are shown for the five B-mode templates described in \refsec{sim_template};
 the points and shaded regions show the mean and variance, respectively.
 The theoretical lensing B spectrum (no foregrounds) is shown as the solid black line.
 These templates are effectively Wiener-filtered; noisier templates consequently have less power.
 The $\hat{B}_{\rm true}$ auto-spectrum recovers the input lensing spectrum as expected while the realistic template $\hat{B}_{\bar{E} ,\phi_{\rm CIB}}$ recovers roughly 20\% of the input spectrum.
 This filtering works as designed and we find that the auto-spectrum of each template corresponds well with the final delensing efficiency.
 From this figure one can already see that the primary factor limiting the delensing efficiency of this analysis is decorrelation in the estimate of $\phi$; see \refsec{discussion} for details.
}
\label{fig:bhat_auto}
\end{center}
\end{figure}

\begin{figure*}
\begin{center}
\includegraphics[width=0.8\textwidth]{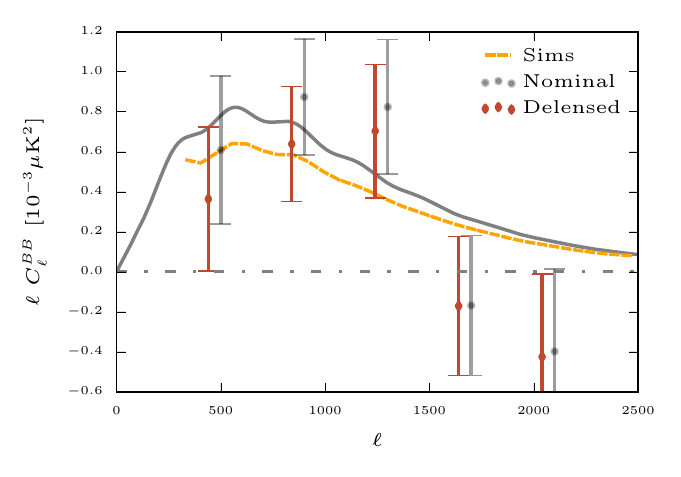}
\caption{
This figure shows the spectrally-combined B-mode bandpowers before (\textcolor{gray}{$\bullet$}) and after delensing (\textcolor{orange}{$\vardiamond$}).
The error bars represent the variance of realistic, noisy simulations.
To guide the eye, the solid line shows the theory spectrum (lensed B modes) and the orange dashed line denotes the expectation value of the delensed spectrum from 100 simulations.
In order to match the data multipole range used in this work, we only plot the simulation results for multipoles $\ell > 300$.
We find that delensing reduces the amplitude of the B-mode spectrum by \effData\%.
}
\label{fig:bandpowers}
\end{center}
\end{figure*}

In order to study the factors that affect delensing efficiency, we create 100 realizations of several different B-mode templates.
We use these different templates to study the delensing efficiency in \refsec{discussion}.
In creating B-mode templates, we can choose whether or not to add a number of non-idealities to the E or $\phi$ maps.

For E maps, variants on three types of simulations are used:
\begin{enumerate}[leftmargin=0.5cm]
 \item{\textbf{``Truth'' maps ($\mathbf{E_{\rm true}}$)}: formed from directly projected Q, U simulated maps with no foregrounds, instrumental or processing effects.}
  \item{\textbf{Filtered, noise-free maps ($\mathbf{\bar{E}_{\rm nf}}$)}: mock-observed, CMB + foreground skies including instrumental and processing effects, but no instrumental noise; $\bar{E}$ maps are derived from the resulting Q and U maps using the filtering process described in Sections~\ref{sec:processing_maps} and \ref{sec:btemplate}.}
  \item{\textbf{Realistic maps ($\mathbf{\bar{E}}$)}: the most realistic simulations, these include all of the steps of the previous section with the addition of instrument noise and the filtering process of \refsec{btemplate}.}
\end{enumerate}
These E maps can then be further masked in 2d harmonic space (e.g., to remove low-$\ell_x$ modes).

For $\phi$ maps, variants on two types of simulations are used:
\begin{enumerate}[leftmargin=0.5cm]
  \item{\textbf{``Truth'' maps ($\mathbf{\phi_{\rm true}}$)}: directly projected CMB lensing potential simulations.}
  \item{\textbf{CIB estimate ($\mathbf{\phi_{\rm CIB}}$)}: simulations of the lensing potential estimated from the CIB, constructed as described in \refsec{sim_pipeline}}
\end{enumerate}

B-mode templates are then constructed from combinations of E and $\phi$ maps from the preceding lists.
For illustration purposes, we show maps from the following five templates in \reffig{templ-maps} and their power spectra in \reffig{bhat_auto}:
\begin{enumerate}[leftmargin=0.5cm]
  \item{\textbf{``Truth'' B modes ($\mathbf{\hat{B}_{\rm true}}$)}: formed from directly projected Q, U simulated maps.}
  \item{\textbf{Ideal estimator template ($\mathbf{\hat{B}}_{E_{\rm true}, \phi_{\rm true}}$)}: template estimated from ideal E and $\phi$ maps.}
  \item{\textbf{Filtered E template ($\mathbf{\hat{B}}_{\bar{E}, \phi_{\rm true}}$)}: template from noisy, filtered E and ideal $\phi$.}

  \item{\textbf{CIB-$\phi$ template ($\mathbf{\hat{B}}_{E_{\rm true}, \phi_{\rm CIB}}$)}: template from ideal E and CIB estimate of $\phi$.}
  \item{\textbf{Realistic template ($\mathbf{\hat{B}}_{\bar{E}, \phi_{\rm CIB}}$)}: template from realistic E and CIB estimate of $\phi$. This template is expected to realistically describe the template constructed from data.}
\end{enumerate}

First, in \reffig{templ-maps} it is visually apparent that the reconstructed templates are recovering much of the structure of $\hat{B}_{\rm true}$.
Second, the B-mode templates are effectively Wiener-filtered; noisier templates consequently have less power.
This is seen clearly in the bottom two template maps shown in \reffig{templ-maps} and the respective average auto-spectra shown in \reffig{bhat_auto}.
Without this filter, extra noise power would be present in the B-mode templates and would cause their auto-spectra to far-exceed that of the lensing B-mode spectrum.

We test the efficacy of our filter on simulations by comparing the auto-spectrum of each B-mode template with the cross-spectrum of the B-mode template with $\hat{B}_{\rm true}$.
In all cases, the mean in $\ell$ of these auto-spectra are $\lesssim 3\%$ larger than the mean of these cross-spectra.
This means that there is very little ``noise" in the templates -- almost all of the power in the B-mode templates is correlated with $\hat{B}_{\rm true}$.
Thus noise in the B-mode templates adds a negligible amount of noise bias to the delensed spectra compared to our statistical error bars.
The delensing efficiency of these templates is discussed further in \refsec{discussion}.

\begin{figure*}
\begin{center}
\includegraphics[width=0.92\textwidth]{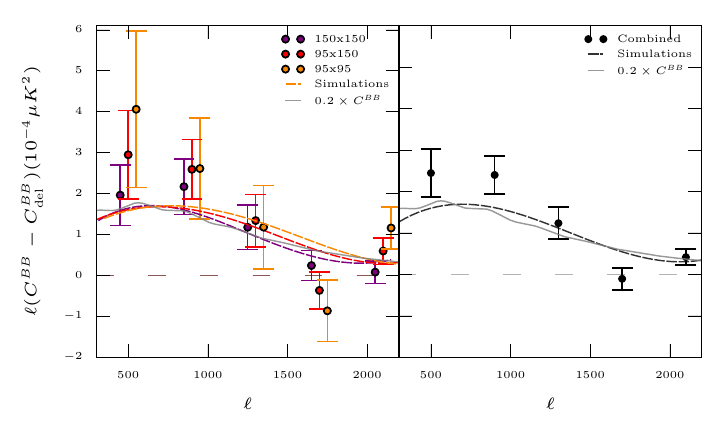}
\caption{This figure presents our main result: the difference between nominal and delensed bandpowers $\Delta C_{\ell}^{BB}$ defined in \refeq{diff}.
The error bars show the variance of this difference in the simulations.
The dashed lines show the (smoothed) average of differenced bandpowers in simulations.
The three differenced auto- and cross-spectra are shown on the left; the minimum variance combined differenced bandpowers are shown on the right.
To guide the eye, we also add a theoretical line that corresponds to $20\%$ of the lensing power.
We use the variance to quantify the significance of the reduction of lensing B power through the delensing process and find that the no-delensing hypothesis is ruled out at \delSignificance$\sigma$.
}
\label{fig:delta_cl}
\end{center}
\end{figure*}

\section{Results}
\label{sec:res}

This section presents the main results of the paper, starting with the expected delensing efficiency from simulations and ending with the delensed \sptpol B-mode power spectrum.

\subsection{Expectation from Simulations}
\label{sec:res-sim}
Before looking at the data, we calculate the expected level of delensing using the simulations described in \refsec{sim}.
The ``realistic template'' $\hat{B}_{\bar{E}, \phi_{\rm CIB}}$ is used to delens the corresponding noisy simulated B maps.
The mean delensed spectrum from 100 simulations is shown in \reffig{bandpowers} by the orange dashed line.
This is the expectation value of the delensed power spectrum.

Using these simulations, we fit the mean bandpowers from the 100 simulations to an \Alens-scaled BB spectrum
for both the nominal and the delensed case.
The \Alens-scaled BB spectrum used here does not include foregrounds even though they are present both in the sims and in the data.
\reftext{We test the foreground modeling in \refsec{fgnd}}.
We find that delensing is expected to reduce the best-fit amplitude from $\Alens = 1.09 \pm 0.29$ to $\Ares = 0.87 \pm 0.28$.
The expected delensing efficiency, calculated as the value of $\alpha$ averaged over these simulations, is
\beq
\langle \alpha \rangle_{\rm sims}\, = 0.23 \pm 0.10.
\eeq

In the limit of the B-mode measurement having zero noise, the fractional reduction in lensing B-mode power through delensing corresponds to the fractional reduction in lensing B-mode sample variance.
In this work, since the variance of the B-mode measurement is dominated by the instrument noise, we do not expect a significant reduction in the variance of the delensed B-mode bandpowers.
This can be seen already from the marginally reduced uncertainty of \Ares compared to \Alens in simulations.

\subsection{Data}
\label{sec:res-data}

The \sptpol B-mode maps described in \refsec{bb_maps} are delensed using the B-mode template described in \refsec{btemplate}.
The nominal and delensed B-mode bandpowers are shown in \reffig{bandpowers}.
It is clear by eye that the delensing process removes some of the B-mode power; that is, delensing is at least partially successful.

To highlight the power removed by delensing, the spectrum difference for the data, as defined in \refeq{diff}, is shown in \reffig{delta_cl} for the spectrally-combined bandpowers as well as the individual frequency band auto- and cross-spectra.
The error bars in this plot show the variance of the spectrum difference for the realistic simulations and the dashed line corresponds to the mean spectrum difference from simulations.
Note that the spectrum difference from the data is consistent with the expectation from simulations.

To quantify the power removed by the delensing process,  following the same approach used for sims in  \refsec{res-sim}, the nominal bandpowers are fit to a lensed B-mode spectrum yielding an amplitude of
\beq
  \Alens = \AlensNom \,.
\eeq
After delensing, the bandpowers are re-fit to the lensed B-mode spectrum; the best-fit amplitude of the residuals is
\beq
  \Ares = \AlensDel \,.
\eeq
This is consistent with the value obtained in simulations of \Ares = $0.87 \pm 0.28$.
The lensing efficiency of the data defined in \refeq{efficiency} is thus $\alpha = 0.\effData\,$, consistent with expectations.

We quantify the significance of our delensing process with the spectrum difference \Dclbb defined in \refeq{diff} and shown in \reffig{delta_cl}.
We use an amplitude $A^{\Dclbb}$ that scales a model \Dclbb spectrum, which we take to be the mean spectrum difference of our realistic noisy simulations.
The amplitude is normalized such that the expectation from simulations is $A^{\Dclbb}=1.$
The model spectra are shown as dashed lines in \reffig{delta_cl}.
We fit for the amplitude which minimizes the $\chi^2$ of the data $\Dclbb$ relative to this model spectrum.
Using 15 bandpowers (5 bins each for the 3 $\Dclbb$ spectra: $95\times 95,95\times 150, 150\times 150$) and a diagonal covariance matrix (as in K15, see there for details) calculated from the variance of our realistic simulations, we find a best-fit amplitude of
\beq
  \label{eqn:Dclbb_data}
  A^{\Delta C_{\ell}^{BB}}=1.18\pm 0.17 \,.
\eeq
Thus the measured \Dclbb spectrum is consistent with the expectation from simulations within $\sim 1\sigma$, and no delensing is ruled out at
\delSignificance$\sigma$.

In order for delensing to improve constraints on PGW B-modes, the delensing process must reduce the variance of the B-mode measurement.
Without delensing, one can simply fit a measured $C_{\ell}^{BB}$ spectrum to a model with primordial and lensing components, however, this process retains the full lensing sample variance.
In contrast, by removing the lensing signal specific to the relevant patch of sky, the delensing process should remove some of the sample variance in the lensing component, thus reducing the variance of the residual bandpowers.
In this analysis, the overall power is reduced, however, as is shown from our simulations, the variance of the residual spectrum is minimally reduced at the level of {$1-3\%$};
this is because the variance is still dominated by noise in the B-mode maps, which is not reduced by delensing.
As mentioned, in the limit of zero map noise, the fractional reduction in the lensing B-mode power corresponds to the fractional reduction in the lensing B-mode variance.
Therefore, as we look forward to future experiments with lower-noise maps, delensing will increase the reduction in variance of the lensing B modes compared to this analysis.

\begin{figure}
\begin{center}
\includegraphics[width=0.48\textwidth]{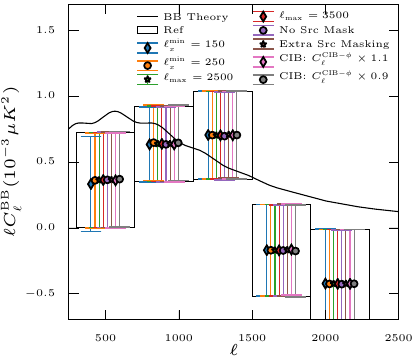}
\caption{\textbf{Systematic tests:}
 This figure shows the delensed data spectrum $C^{del}_{b,\,{\rm sys, data}}$ for each of the changes to the B-mode template described in \refsec{sys}.
 The error bars are from the variance of the delensed spectrum from simulations $C^{del}_{b,\,{\rm sys, i}}$.
 `Ref' denote the reference delensed bandpowers using the fiducial B-mode template we utilized for the main result.
 In all cases the delensed spectrum changes by less than 5\% of the statistical uncertainty; the delensed spectrum is very insensitive to these analysis choices.
 }
\label{fig:sys}
\end{center}
\end{figure}

\subsubsection{\reftext{Robustness Against Foreground Modeling}}
\label{sec:fgnd}

\reftext{
In this section, we test the robustness of the delensing result against our foreground modeling assumptions.
Foregrounds could enter this analysis in two places: in the E-mode maps that are used to form the B-mode templates, and in the B-mode maps that are used to calculate bandpowers.
The only way foregrounds could contribute to the observed delensing result is if foreground components in the E-mode maps couple to the lensing estimator in a way that produces modes in the lensing B-mode template which correlate with the measured B-mode maps (either the lensing or foreground component).
Otherwise, foregrounds in the E-mode maps will merely add noise to the B-mode template, while foregrounds in the B-mode maps will be unaffected by the delensing operation and thus be removed when the nominal and delensed bandpowers are differenced.
In the current analysis, we expect the foreground contamination of the delensing result through the above mechanism to be vanishingly small relative to the noise levels.
Indeed, the foreground power in our E-mode and B-mode maps is already small (see C14 and K15) and the non-Gaussian contributions at the angular scales of interest should be even smaller.}

\reftext{
We test the above expectation by fitting the data to models including foreground components.
We use the foreground model from K15, which includes four nuisance parameters for foreground emission from polarized point-sources and dust; see K15 for details.
The final model has six parameters: $A_{\rm lens}$, $r,$ and four foreground parameters.
We then fit the nominal and delensed bandpowers with this model using the Markov chain Monte Carlo method provided by the CosmoMC\footnote{http://cosmologist.info/cosmomc/} software package \citep{lewis02b}.
To study the impact of dust, we run two separate chains with different priors on the dust amplitude $A_{\rm dust}$: 
the nominal prior is a Gaussian centered on 1.0 with $\sigma = 0.3$, and the second prior is flat between 0.7 and 1.3, where $A_{\rm dust}=1$ is the normalized dust amplitude for the 3 auto- and cross- spectral bands in this analysis. 
The resulting delensing efficiency for the six-parameter fits with Gaussian and flat prior models are $31\%$ and $28\%$, respectively; 
these are consistent with the efficiency of $28\%$ reported above.
The dust amplitude constraint is entirely dominated by the prior and does not change pre- and post-delensing in either case.
The delensing efficiencies are consistent regardless of how the foregrounds are modeled in the fit;
therefore, we conclude that the measured delensing efficiency is unaffected by foregrounds in this analysis.
}

\section{Systematics, Data Consistency, and Significance Tests}
\label{sec:sys-null-sign}
In this section, we describe several tests we performed to ensure the delensing results are both robust against systematic errors in the data and also statistically significant.
For each test, we change one aspect of the B-mode template construction pipeline and recompute the delensed spectrum ($C^{del}_{\rm sys/null}$).
We then calculate the difference between this spectrum and the spectrum from a reference delensing analysis,
defined as the following:
\beq
\label{eqn:delta_test2}
  \Delta C^{del}_{b,\,{\rm sys/null}} = C^{del}_{b,\,{\rm sys/null}} - C_{b,\,{\rm ref}}\,,
\eeq
where the reference spectrum $C_{b,\,{\rm ref}}$ is defined differently for the systematics tests or null tests and is specified below.
These spectra are calculated for the data ($\Delta C^{del}_{b,\,{\rm sys},\,{\rm data}}$) as well as each simulation ($\Delta C^{del}_{b,\,{\rm sys},\,i}$).

\subsection{Systematics Tests}
\label{sec:sys}
Systematics tests check the robustness of the delensing result to analysis choices in constructing the B-mode template.
For each systematics test, we change one analysis choice in the B-mode template construction, and recompute the delensed spectrum ($C^{del}_{\rm sys}$).

The following systematics tests are performed:
\begin{enumerate}[leftmargin=0.5cm]
  \item{\textbf{E map $\bl_x$ cut}: in our baseline analysis we cut all modes with $|\ell_x| < 200$ from the E-mode maps used to construct the B-mode template.
 In this test, we adjust that cut from the reference value $|\ell_x| < 200$ to $|\ell_x| < 150$ and $|\ell_x| < 250$.
}
  \item{\textbf{E map $\bl_{max}$ cut}: We change the maximum value of $\ell$ modes in the E-mode CMB maps used to reconstruct the B-template from the nominal value of $\ell_{max}=3000$ to $\ell_{max}=2500$ and $\ell_{max}=3500$.}

  \item{\textbf{Point source masking}: In the baseline analysis, we mask all the sources with flux $S_{150}>50$ mJy.
    We repeat the same analysis without masking any source (``No Src Mask'') and by masking more aggressively every source with flux $S_{150}>6.4$ mJy (``Extra Src Masking'') in the E-mode maps that are used to construct the B-mode template.
}

\item{\textbf{CIB-$\phi$ correlation}:
We simulate CIB realizations as in \refsec{sim_pipeline} using the nominal $C_{\ell}^{\rm{CIB}-\phi}$ correlation but analyze them assuming a rescaled correlation $A \cdot C_{\ell}^{\rm{CIB}-\phi}$.
The result for $C^{del}_{\rm sys}$ shows how much using an incorrect Wiener filter for the CIB map changes the delensed spectrum.
With the $C_{b,\,{\rm ref}}$ fixed to the nominal $C_{\ell}^{\rm{CIB}-\phi}$, we compute the $\chi^2$ of the spectrum difference
$\Delta C^{del}_{b,\,{\rm sys}}$ between simulations and data for $A=[0.9,1.1]$. This corresponds to a $10\%$ uncertainties in the amplitude of $C_{\ell}^{\rm{CIB}-\phi}$ following the results of \citet{planck2013XVIII}.
}
\end{enumerate}
For these tests, we use the nominal delensed bandpowers for either data ($C^{del}_{b,\,{\rm data}}$) or simulations ($C^{del}_{b,\,i}$) as the reference spectrum $C_{b,\,{\rm ref}}$ in \refeq{delta_test2}.
If a systematic were present in the data, such an analysis change would cause a larger spectrum difference than expected from the simulations.

\begin{figure}
\begin{center}
\includegraphics[width=0.47\textwidth]{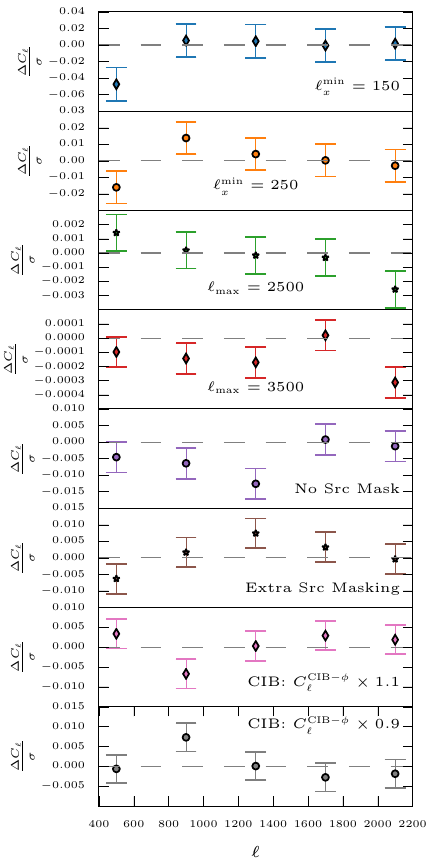}
\caption{\textbf{Systematic tests:}
  This figure shows the spectrum difference $\Delta C^{del}_{b,\,{\rm sys},data}$ for each of the changes to the B-mode template described in \refsec{sys}, divided by the $1\sigma$ statistical uncertainty of the baseline delensed data spectrum ($C^{del}_{b,\,{\rm data}}$).
  The error bars are calculated from the variance in simulations of each consistency test, i.e the variance of ($\Delta C^{del}_{b,\,{\rm sys},\,i}$).
  We find the spectrum difference to be compatible with the changes observed in simulations, and small (less than 5\%) compared to the statistical uncertainty.
}
\label{fig:sys_diff}
\end{center}
\end{figure}

The results of these tests are presented in two figures: \reffig{sys} shows the systematic delensed spectra $C^{del}_{\rm sys, data}$, and \reffig{sys_diff} shows the spectrum difference for each test normalized by the statistical uncertainty.
We find that the systematic delensed spectra change negligibly (less than 5\% of the statistical uncertainty, see \reffig{sys_diff}) compared to the baseline analysis for all the systematics tests performed.
Furthermore, as seen in \reffig{sys_diff}, these changes are compatible with the changes we observe in the simulations.

Spurious correlations between the B-mode data and the lensing B-mode template could in principle bias the delensing result presented here. Possible sources of such correlations are galactic foregrounds or E-to-B leakage. However we expect any such bias to be negligible in this noise-dominated analysis (see additional discussion in section 4B of \citealt{sherwin15}).
Indeed such a bias would be present if a contaminating component in the E map propagated through the lensing estimator into the B template in a way that correlated with the measured B map.
Any contamination from E-to-B leakage and foregrounds through this mechanism should be very small.
Additionally, in this analysis we are protected from possible spurious correlations between the lensing potential and the CMB maps because $\phi$ is estimated from the CIB as opposed to the CMB itself.
While negligible now, biases of this nature could become important in the future.

Therefore, we conclude that our measured delensing result is not contaminated by these systematics.

\subsection{Null / Consistency Tests}
\label{sec:null}

\begin{figure}
\begin{center}
\includegraphics[width=0.4\textwidth]{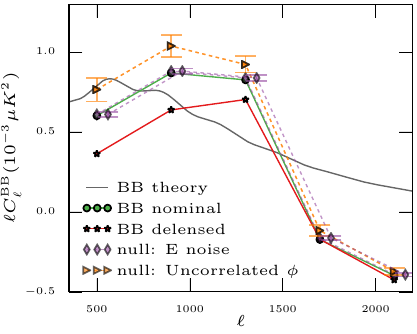}
\caption{\textbf{Null tests:}
This figure shows the nominal (green) and delensed (red) bandpowers as well as the ``E noise'' (purple) and ``Uncorrelated $\phi$'' (orange) bandpowers.
The ``E noise'' bandpowers almost completely overlap with the nominal bandpowers and are slightly shifted.
For these tests, the data points show the average null-delensed bandpowers and the errorbars show $\sigma_{b,\, {\rm null}}$ as defined in \refsec{null}.
Note that when uncorrelated templates are used, the delensing process just adds noise thus \textit{increasing} the measured power.
The null tests are quantified in \reftab{systematics}.
}
\label{fig:null}
\end{center}
\end{figure}

Null / consistency tests check the significance of our result by delensing using a null template in place of the nominal B-mode template.
The null templates will still have power, however, this power is expected to be uncorrelated with the B modes on the sky (or in the simulations);
thus delensing with the null template will \textit{add} power on average to the BB bandpowers.
We perform two null tests: E noise and uncorrelated $\phi$.

To understand these tests, we pose the following null hypothesis:
``The change in bandpowers after delensing arises from a statistical fluctuation of the E-mode noise or the CIB map, ''
and ask how likely it is that the delensed bandpowers we see in our main result are due to such statistical fluctuations.

We form null templates for each test as follows:
\begin{enumerate}[leftmargin=0.5cm]
  \item{\textbf{E noise}: in the baseline analysis on data, replace the E map with 100 noise realizations.
     }
  \item{\textbf{Uncorrelated $\mathbf{\phi}$}: in the baseline analysis on data, replace the $\hat{\phi}_{\rm CIB}$ with 100 simulations.
     }
\end{enumerate}
The null-delensed spectra $C^{del}_{b,\,{\rm null},\,{\rm data}}$ are shown in \reffig{null}.

The ``E noise'' B-mode template has little power, and hence has little effect on the BB bandpowers.
This is because the \sptpol E maps are signal-dominated, thus the pure-noise E maps simply have much less power;
additionally, the filtering described in \refsec{btemplate} heavily suppresses noisy modes.
The ``Uncorrelated $\phi$'' B-mode template has power at a level similar to the default B-mode template, however, the modes are uncorrelated with the B modes on the sky.
Thus delensing with this template adds significant power to the BB bandpowers.

To quantify this statement, we calculate how consistent delensed data bandpowers are with the null distribution using a $\chi^2$ test as
\beq
\label{eqn:chisq_null2}
  \chi^2_{\rm null} = \sum_b{ \frac{\left[(C^{del}_{b,\,{\rm data}} - C^{BB}_{b,\,{\rm data}}) - \left<\Delta C^{del}_{b,\,{\rm null}\,,i}\right> \right]^2} { \sigma_{b,\,{\rm null}}^2} } \\
\eeq
Here $\left<\Delta C^{del}_{b,\,{\rm null}\,,i}\right>$ is the expectation value of the spectrum difference between the null-template delensed bandpowers and $C^{BB}_{b,{\rm data}}$,
and $\sigma_{b,\,{\rm null}}^2$ is the variance of the null distribution $\Delta C^{del}_{b,\,{\rm null}\,,i}$.
We find large values of $\chi^2_{\rm null}$ (see \reftab{systematics}) showing that the delensing result is inconsistent with the null hypothesis.

\begin{table}
\caption{Null tests and their significances}
\centering
\begin{tabular}{|c |c|}
\hline\hline
Null test & $\chi^2$   \\
\hline
 $E:$ noise only          &  932.9   \\
  Uncorrelated $\phi$       &  174.7  \\
\hline
\end{tabular}
\vspace*{2mm}
\tablecomments{\textbf{Null tests:}
  For the two null tests described \refsec{null}, we list the $\chi^2$ for 15 degrees of freedom calculated from comparing the nominal spectrum difference relative to the distribution of null-delensed spectrum difference, as defined in \refeq{chisq_null2}.
  We find that the nominal spectrum difference is very \textit{inconsistent} with the distribution of null-delensed spectrum difference; this rules out the null hypothesis that the observed delensed bandpowers result from a statistical noise fluctuation in either E or $\phi$.
  \vspace*{5mm}
}
\label{tab:systematics}
\vspace{0.1cm}
\end{table}

\section{Understanding Delensing Efficiency}
\label{sec:discussion}

In this section, we use analytic tools and our simulations to study how different factors affect delensing efficiency.
Additionally, future-experiment planning requires tools that accurately predict the level of delensing residuals given experiment configurations and survey strategy.
Analytic tools that have been developed so far usually assume ideal experiment inputs, for example, isotropic noise and complete, uniform sky coverage.
In this section, we first show that standard analytic tools are lacking in accounting for realistic non-idealities in experiment inputs.
Then, with the suite of simulations we built for this work, we show how various aspects of experiment non-idealities affect future experiments' delensing efficiencies.
Because the goal of this section is to understand the factors that affect delensing efficiency and develop intuition on aspects in forecasting tools that need to be modeled with better care going forward,
a complete forecast of future experiments' delensing efficiency given realistic experiment constraints is left for follow-up work.

We have two tools at our disposal for this investigation: analytic calculations and simulations.
The analytic calculations here use the idealized expression for delensed residuals, written as the full-sky equivalent of \refeq{clbb_res}.
In the simulations, we have the ability to separate out different factors such as noise, decorrelation in $\phi$, and modes that have been suppressed or removed.
Noise in the B maps merely adds variance to these tests, so we eliminate noise variance in the B-mode power spectrum by calculating $C_{\ell}^{BB}$ from noiseless simulated B maps.
These noiseless B-mode maps are then delensed using B-mode templates constructed from various combinations of E and $\phi$.
The delensing efficiency $\alpha$ for each case is quantified using \refeq{efficiency}.
We obtain \Alens from simulations by fitting the mean of the nominal B-mode spectra to a two-component, one-parameter model comprised of the input lensing B-mode spectrum and the fixed input foreground spectrum, where \Alens scales the lensing component.
We obtain \Ares analogously.
The foreground spectrum is computed with parameters described in \refsec{sim}.

This fitting procedure differs from the method in \refsec{res} by including a fixed foreground component.
Ideally, we would have simply used sims without foregrounds, however, that would be computationally expensive since the foregrounds are added before creating mock time ordered data for each sim (see \refsec{sim}).
Instead, because the foreground contributions to these sims are known exactly we choose to account for their contribution to the fit as described above.
A detailed investigation of the effects of foregrounds is outside the scope of this work.

We emphasize that the results presented in this section are self-consistent but cannot be directly compared to those from \refsec{res} for three reasons:
the spectra from \refsec{res} use five bins whereas here we use much smaller bins ($\delta_{\ell} = 100$),
the data and sims from \refsec{res} contain noise, whereas here we use noiseless B maps,
and the foregrounds are treated differently.


\subsection{Connecting to Analytic Expectations}

To show how the standard analytic forecast tools are lacking in capturing experiment input non-idealities,
we calculate the delensing efficiency expected from the idealized analytic expression for delensed residuals
and compare that to the one achieved from our realistic simulations.

To derive the delensing efficiency from the analytic expression, we model $N_{\ell}^{EE}$ as a 9~$\mu {\rm K}$-arcmin white noise spectrum, include a 1$^{\prime}$ beam, and choose $\rho_{\ell}$ as described in \refsec{cib} and shown in \reffig{CIB}.
To approximate cuts made in the data, we exclude E-modes $|\pmb{\ell}| < 200$ and $\phi$ modes $|\pmb{L}| < 150$.
This is less stringent than the E-mode cut in our data and in our simulations, where we remove modes with $|\ell_x| < 200$.

In order to to extract the delensing efficiency, we fit the delensed theory curve to a scaled lensing theory spectrum, similar to getting \Ares from the delensed bandpowers in simulations.
With the above inputs ($E^{9\mu {\rm K}^{\prime}}_{\rm true, \ell_{min} = 200}$, $\phi_{\rm CIB}$), \refeq{clbb_res} predicts a delensing efficiency of 27\% (4th entry in the analytic $\alpha$ column in \reftab{eff_ebarpcib}).
When we include almost all $\phi_{\rm CIB}$ modes ($L_{\rm min} = 20$) and use a noiseless $E_{\rm true}$, the delensing efficiency improves to 31.5\%  (1st entry in the analytic $\alpha$ column in \reftab{eff_ebarpcib}).
The difference between these two analytic cases comes mostly from the $L$ cut in $\phi$; when this cut is restored, the efficiency drops back to 28.5\%.

The expected delensing efficiency from the mean of 100 realistic simulations using $\hat{B}_{\bar{E}, \phi_{\rm CIB}}$ is 19.7\%;\footnote{
  This value is different from the value of $\langle \alpha \rangle_{\rm sims} = 0.23 \pm 0.10$ obtained in \refsec{res-sim} due to the different binning, noiseless B maps, and treatment of foregrounds in the fit.
  The value obtained here is consistent with the rest of this section and hence useful for understanding delensing efficiency, but is not directly comparable to the data.}
whereas if one uses the analytic expression to calculate the delensing residuals, one would get delensing efficiency of 27\%.
To account for the differences between these two results, we first start with comparing the delensing efficiencies of B-mode templates formed from
the ideal, no mode-loss $E_{\rm true}$ in our simulations to those from analytic expressions.
Then we inject imperfections to $E_{\rm true}$ to approximate it to $\bar{E}$, and calculate the delensing efficiency
for each injected imperfection.
The results are tabulated in \reftab{eff_ebarpcib}.

The $\sim 7\%$ difference in efficiency between the analytic expression and our simulations is accounted for and explained as follows:
(1)
comparing delensing efficiencies from B-mode templates made from ($E^{9\mu {\rm K}^{\prime}}_{\rm true}$, $\phi_{\rm CIB}$)
with $\ell_{\rm x min} = 200$ and with $\ell_{\rm min} = 200$ cut, one can see that $\sim1.5\%$
of the 7\% difference comes from
implementing a $\ell$ cut instead of a $\ell_x$ cut;
(2) comparing delensing efficiencies from B-mode templates made from $E_{\rm true, \ell_{min} = 200}$ and $\phi_{\rm CIB}$,
but with two levels of noise, 9 and 12~$\mu {\rm K}$-arcmin, on the E-mode map, we see that $\sim1\%$ of the  7\% difference comes from
non-white noise that is not well-approximated by a simple 9~$\mu {\rm K}$-arcmin $N_{\ell}$;
(3) comparing between ($E_{\rm true}$, $\phi_{\rm CIB}$) with nominal mask and with extended mask (ext. mask),
one can see that the delensing efficiency is degraded by 0.5\% when one does not include E/$\phi$ modes outside of the B-mode survey boundary when forming the B-mode template.
The rest of the difference ($\sim4\%$) are from higher-order lensing effects, ``noise" from the B-mode template
(part of the B-mode template that is not correlated with lensing B modes as discussed in \refsec{sim_pipeline}), and effects beyond the scope of this paper's investigation.
While these effects are not dominant in the current noise regime, they will become proportionately more significant in making forecasts for low-noise regimes.

Thus while the differences between the analytic and simulation-based efficiencies are understood,
it is clear that a naive application of the standard delensing forecasting over-predicts the achieved delensing efficiency.
This highlights the need to use realistic simulations or better analytic tools that incorporate realistic experiment non-idealities for forecasting delensing efficiencies for experiment planning.

\begin{table}
\caption{Delensing efficiency: \\ comparing simulations to analytic predictions}
\centering
\begin{tabular}{|l | c | c |}
\hline\hline
B-mode template inputs & $\alpha$: Simulations  & $\alpha$: Analytic \\
\hline
\vspace{-0.28cm}&  & \\

$E_{\rm true}$, $\phi_{{\rm CIB}, L_{\rm min} = 20}$		         		& N.A.      & 31.5\%    \\
\vspace{-0.28cm}&  & \\
$E_{\rm true}$, $\phi_{\rm CIB}$ (ext. mask)				      		 & 27.1\% & N.A.      \\
\vspace{-0.28cm}&  & \\
$E_{\rm true}$, $\phi_{\rm CIB}$ 		     						& 26.6\% & 28.5\%    \\
\vspace{-0.28cm}&  & \\
$E^{9\mu {\rm K}^{\prime}}_{\rm true, \ell_{min} = 200}$, $\phi_{\rm CIB}$       & 22.2\% & 27.0\%  \\
\vspace{-0.28cm}&  & \\
$E^{9\mu {\rm K}^{\prime}}_{\rm true, \ell_{x min} = 200}$, $\phi_{\rm CIB}$    & 20.7\% & N.A.     \\
\vspace{-0.28cm}&  & \\
$E^{12\mu {\rm K}^{\prime}}_{\rm true, \ell_{x min} = 200}$, $\phi_{\rm CIB}$  & 18.7\%  & N.A.      \\
\vspace{-0.28cm}&  & \\
$\bar{E}$, $\phi_{\rm CIB}$  						     			 & 19.7\%   & N.A.      \\
\hline
\end{tabular}
\vspace{0.1cm}
\tablecomments{
This table decomposes the factors affecting the delensing efficiency of the realistic ($\bar{E}$, $\phi_{\rm CIB}$) simulations,
and shows the analytic prediction from \refeq{clbb_res} where applicable.
Starting from $E_{\rm true}$, we inject imperfections that capture the features of $\bar{E}$ that account for the disconnect between the analytic and simulated delensing efficiency.
The table progresses from ideal at the top to realistic at the bottom.
Note, unless otherwise stated, all of the $\phi_{\rm CIB}$ cases have $L_{\rm min}$ cut of $L=150$;
$E_{\rm true}$ is noiseless and has no mode-loss;
$\bar{E}$ is the realistic simulation with instrument noise and effects of filtering.
The efficiencies are calculated by fitting the residual B power to a scaled lensing B-mode power spectrum with fixed foreground spectrum.
Effects from $\ell_x$ cuts and from masking for future experiments are further discussed in \refsec{effsim}.
\vspace{0.3cm}
}
\label{tab:eff_ebarpcib}
\end{table}

\subsection{Understanding Efficiency from Simulations}
\label{sec:effsim}
In this section, we use the simulation suite we have built for this work to study how the delensing efficiency depends on various input factors.
As outlined in \refsec{sim_template}, B-mode templates can be made using various E-mode map inputs and $\phi$ map inputs.
We classify the factors entering \refeq{clbb_res} that determine the delensing efficiency into three categories (and sub-groups within each category) as follows:
\begin{enumerate}[leftmargin=0.5cm]
  \item{Factors contributing to noise terms in the Wiener filter:}
    \begin{itemize}[leftmargin=0.5cm]
      \item{Decorrelation in $\hat{\phi}$}
      \item{Noise in $\bar{E}$}
    \end{itemize}
   \item{Factors contributing to signal terms:}
      \begin{itemize}[leftmargin=0.5cm]
     	 \item{Low $L$ cut in $\hat{\phi}^{\rm CIB}$}
     	 \item{Missing modes in $\bar{E}$}
      \end{itemize}
  \item{Other factors that are not explicit in \refeq{clbb_res}:}
    \begin{itemize}[leftmargin=0.5cm]
   	 \item{Masking/apodization}
   	 \item{Beyond $\mathcal{O}(\phi)$ lensing}
    \end{itemize}

\end{enumerate}

\begin{figure}
\begin{center}
\includegraphics[width=0.48\textwidth]{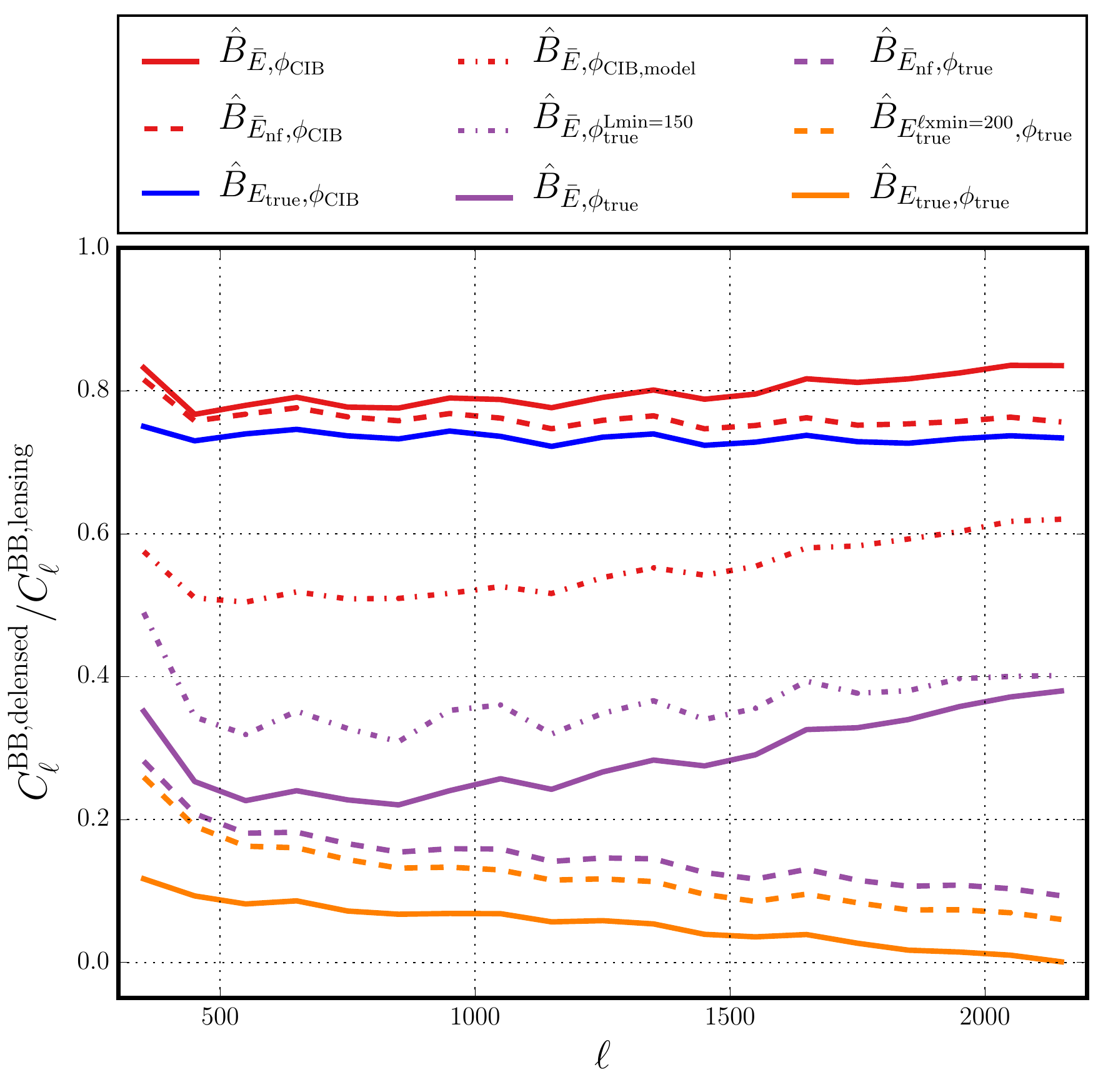}
\caption{Fraction of delensed BB residuals for a few representative cases in simulations.
There is no delensing if the fraction is one; and perfect delensing if the fraction is zero.
The cases drawn with solid lines match the colors of the B-mode template auto-spectra shown in \reffig{bhat_auto}.
The dashed line variant of the same color represents a modification to the E-mode input to the solid line case.
Similarly, the dash-dot line variant represents a modification of the $\phi$ input.
Holding $\phi_{\rm CIB}$ fixed and improving the E-mode input does not drastically reduce the delensed residuals (e.g. from solid red to solid blue);
on the other hand, when $\bar{E}$ is held fixed, using $\phi_{\rm true}$ or $\phi_{\rm CIB, model}$ to form the B-mode template significantly reduces the delensed residuals (e.g. from solid red to dash-dot red, and then to solid/dash-dot purple).
In the noiseless E and $\phi$ cases ($E_{\rm true}$, $\phi_{\rm true}$), not excluding $\ell_x < 200$ modes reduces the delensed residuals by a factor of $\sim2$ (from dashed orange to solid orange).
}
\label{fig:deleff}
\end{center}
\end{figure}

\reftab{eff_future} summarizes the delensing efficiencies for the cases discussed in the following sections,
while \reffig{deleff} shows the scale-dependence of the fractional delensed BB residuals for a few representative cases.

\begin{table}
\caption{Delensing efficiency: intuition for future improvements}
\centering
\begin{tabular}{|l | c |}
\hline
B-mode template inputs &  $\alpha$ \\
\hline
\vspace{-0.25cm}& \\
$E_{\rm true}$, $\phi_{\rm true}$ (extended apod)   &  98.5\% \\
\vspace{-0.25cm}&   \\
$E_{\rm true}$, $\phi_{\rm true}$                       & 95.0\%\\
\vspace{-0.25cm}&   \\
$E_{\rm true}^{\ell_{\rm xmin}=200}$, $\phi_{\rm true}$ & 90.0\%  \\
\vspace{-0.25cm}&   \\
$\bar{E}_{\rm nf}$, $\phi_{\rm true}$ 			& 86.5\% \\
\vspace{-0.25cm}&   \\
$\bar{E}$, $\phi_{\rm true}$ 				& 73.0\% \\
\vspace{-0.25cm}&   \\
$\bar{E}$, $\phi_{\rm true}^{L_{\rm min}=150}$ 		& 64.0\% \\
\vspace{-0.25cm}&   \\
$\bar{E}$, $\phi_{\rm CIB, model} $ 				& 45.0\% \\
\vspace{-0.25cm}&   \\
$\bar{E}$, $\phi_{\rm CIB, model}^{L_{\rm min} =150} $ 	& 39.5\% \\
\vspace{-0.25cm}&   \\
$E_{\rm true}, \phi_{\rm CIB}$                          & 26.5\% \\
\vspace{-0.25cm}&   \\
$\bar{E}_{\rm nf}, \phi_{\rm CIB}$                      & 24.0\% \\
\vspace{-0.25cm}&   \\
$\bar{E}, \phi_{\rm CIB}$                               & 20.0\% \\
\hline
\end{tabular}
\vspace{0.2cm}
\tablecomments{
Delensing efficiencies $\alpha$ (rounded to closest 0.5\%) computed using our simulations.
These are used to understand how much various imperfections in the input E and $\phi$ maps will matter for future experiments.
We order the case starting with ideal (top) and add non-idealities as we proceed down the table.
While this one number does not fully capture the shape of the delensed power as a function of multipole $\ell$, these do capture the relative weights of each imperfection.
To address this, \reffig{deleff} shows the fractional delensed residuals as a function of $\ell$.
\vspace{0.2cm}
}
\label{tab:eff_future}
\end{table}

\subsubsection{Efficiency loss from decorrelation in $\hat{\phi}$}
\label{sec:phidecorr}
The fidelity of our estimate $\hat{\phi}$ of the lensing potential is quantified by the CIB-$\phi$ correlation $\rho_{\ell}$.
This correlation is how we parametrize the effective noise in $\hat{\phi}$.
Several sources contribute to the decorrelation between CIB and $\phi$ including the difference in redshift overlap between star-forming dusty galaxies and $\phi$, foreground contamination, and instrument noise, and other factors.
A correlation coefficient of $\rho_{\ell} = 1$ corresponds to the noise-free limit of $\hat{\phi}$ and is studied by replacing $\hat{\phi}^{\rm{CIB}}$ with the input map  $\phi_{\rm true}$ for each simulation.

In the main analysis presented in \refsec{res}, we find that in this work the delensing efficiency is limited primarily by decorrelation in the $\phi$ tracer.
To see this, instead of constructing the simulated baseline B-mode template $\hat{B}_{\bar{E}, \phi_{\rm CIB}}$,
we replace either the E or $\phi$ map with the ideal map to produce $\hat{B}_{E_{\rm true}, \phi_{\rm CIB}}$ and $\hat{B}_{\bar{E}, \phi_{\rm true}}$.
Replacing the baseline $\hat{B}_{\bar{E}, \phi_{\rm CIB}}$ with either $\hat{B}_{E_{\rm true}, \phi_{\rm CIB}}$ or $\hat{B}_{\bar{E}, \phi_{\rm true}}$ improves the efficiency from 20\% to 26.5\% or 73\%, respectively.
Clearly, the non-idealities in $\hat{\phi}$ cause a much larger suppression of our delensing efficiency.

It is useful to ask how much our achieved delensing efficiency could be improved if we had a CIB map that was more than 65\% correlated with $\phi$.
The CIB-$\phi$ correlation in~\cite{planck2013XVIII}, for example, is shown to be $\sim80\%$ and is approximately flat from degree to arcminute angular scales.
As a test case, we build a CIB-$\phi$ correlation $\rho_{\ell}$ that peaks at 85\% at $L \sim 50$ and follows the shape of the correlation curves shown in Figure 13 of~\cite{planck2013XVIII}; the average of this $\rho_{\ell}$ across $\ell$-multipoles is $\sim0.8$.
Using this  $\rho_{\ell}$ with no cut in $L$, we generate simulated lensing potential maps denoted $\phi_{\rm CIB, model}$ and form B-mode templates using this and our realistic $\bar{E}$.
Delensing with this B-mode template, we obtain an average delensing efficiency of 45\%.
Thus CIB maps that are more correlated with the CMB lensing potential can potentially improve the achieved delensing efficiency by more than a factor of two.

\subsubsection{Efficiency loss from noise in E}
At the current level of noise in $\hat{\phi}^{\rm{CIB}}$, the noise in the E maps has little effect;
indeed when $\bar{E}$ is replaced with the noise-free version $\bar{E}_{\rm nf}$ (fixing $\hat{\phi}^{\rm{CIB}}$), the efficiency increases only from 20\% to 24\%.
This small improvement comes mostly at high-$\ell$ where E noise is dominant, as seen by comparing the solid red line to the dashed red line in \reffig{deleff}.

As estimates of $\phi$ improve, however, E-map noise will become an important factor for delensing.
To study this, we construct two B-mode templates with the same perfect noiseless $\phi_{\rm true}$ but two different E realizations:
(1) a realistic, noisy $\bar{E}$ and
(2) a filtered, noise-free $\bar{E}_{\rm nf}$.
The resulting improvements to the delensing efficiency $\alpha$ are again scale-dependent;
the efficiency improves marginally on large scales ($< 5\%$ at $\ell < 500$), but significantly at small scales ($> 25\%$ at $\ell > 2000$),
as can be seen by comparing the solid purple to the dashed purple line in \reffig{deleff}.
This is because \sptpol E maps are signal-dominated at large scales up to $\ell < 1700$ (see C15).

\subsubsection{Efficiency loss from mode-exclusion in $\hat{\phi}$}
In the fiducial analysis, large-scale modes ($L < 150$) are excluded from $\hat{\phi}^{\rm{CIB}}$ due to galactic dust contamination.
At the current level decorrelation between the $\phi$ tracer and $\phi_{\rm true}$, removing these modes has little effect.
However, in the lower-noise limit for the lensing tracer of $\phi$ these modes can matter.

We study the effect on delensing efficiency due to this mode-cut for both the $\phi_{\rm CIB, model}$ and the $\phi_{\rm true}$ case, while fixing the E-mode input to the realistic $\bar{E}$.
When we form a B-mode template using $\phi_{\rm CIB, model}$ as described in \refsec{phidecorr} with and without $L < 150$ modes,
the delensing efficiency $\alpha$ are 39.5\% and 45\% respectively.
Similarly, for the hypothetical noiseless tracer case ($\phi_{\rm true}$; $\rho_{\ell} = 1$), we compare the delensing efficiency of B-mode template formed using $\phi_{\rm true}$ with and without modes $L < 150$ ($\bar{E}$).
We find that when $L < 150$ modes are included in $\phi_{\rm true}$, $\alpha$ increases from 64\% to 73\%.

We note, however, $L < 150 ~ \phi$ modes contribute less to the lensing B-modes at degree scale, where the predicted PGW signal peaks, than to the range we are testing in this work (see e.g. Figure 2 in \citealt{simard:2015}).
Thus recovering $\hat{\phi}^{\rm{CIB}}$ modes at $L < 150$ might be less critical for PGW searches.
Furthermore, with the advent of low-noise CMB experiments, $\phi$ reconstructed using CMB will soon become a higher signal-to-noise tracer of $\phi_{\rm true}$ than the CIB.
In the case of CMB-reconstructed $\phi$, the large angular scales in $\phi$ (e.g. $L < 150$) are sourced mainly from intermediate scales in the CMB.
The foreground contamination on these scales in the CMB is expected to be lower than at very large or very small scales, particularly in polarization.
Therefore, we expect future CMB-based $\phi$ estimates not to require a low-$L$ cut.

\subsubsection{Efficiency loss from missing modes in E}
Modes in the E maps are suppressed or otherwise mis-estimated as a result of timestream filtering, the instrumental beam, masking, and other factors.
As noted earlier, these modes are removed primarily from the scan-synchronous direction approximately aligned with $\ell_x$.
The effects of removing these modes are included in our simulations by mock-observing and filtering the ``truth'' simulated maps.
We study the effect of these missing modes by comparing the delensing efficiency between the un-filtered $E_{\rm true}$ and the filtered $\bar{E}_{\rm nf}$.

At the current noise levels in $\phi$, these missing modes have little effect; indeed replacing $\bar{E}_{\rm nf}$ with $E_{\rm true}$ improves $\alpha$ by only a few percent (see the ninth and tenth entries of \reftab{eff_future}).
In the low-noise limit of E and $\phi$ when the delensing efficiency is high, however, these missing modes can matter significantly.
First, we demonstrate that the effect of filtering (i.e., the difference between $\bar{E}_{\rm nf}$ and $E_{\rm true}$) is well approximated by simply removing modes in the E map.
To do so we construct a B-mode template from ($E_{\rm true}$, $\phi_{\rm true}$) but apply the $\ell$-mask used on the data, removing all E modes with $|\ell_x|<200$ and the resulting delensing efficiency is within a few percent of that from ($\bar{E}_{\rm nf}$, $\phi_{\rm true}$).
Given that, we can extract the effect of missing modes in the E map on the delensing efficiency by comparing the delensing efficiencies of B-mode templates constructed from ($E_{\rm true}$, $\phi_{\rm true}$) and ($E_{\rm true}^{\ell_{\rm xmin}=200}$, $\phi_{\rm true}$).
We see that by including all the modes in the E map, we improve $\alpha$ from 90\% to 95\%.
Looking at this from the delensed residuals perspective, one can see by comparing the solid and dashed orange lines of \reffig{deleff}, when using $E_{\rm true}$ in place of $E_{\rm true}^{\ell_{\rm xmin}=200}$ the delensed residuals are reduced by $\sim5\%$ at all scales in this very low delensed residuals regime.
Thus as measurement noise decreases, it will become proportionately more important to recover as many modes as possible.

\subsubsection{Remaining factors}
Even if the B-template is formed by noiseless and un-filtered fields $E_{\rm true}$ and $\phi_{\rm true}$,
reproducing the conditions of the analytic expression, the delensing efficiencies of the two approaches still differ by $\sim 5\%$.
We investigate two possible causes for this difference: the contribution of E-modes outside our apodization mask to B-modes inside the mask and the impact of higher-order lensing beyond $\mathcal{O}(\phi)$.

\begin{enumerate}[leftmargin=0.5cm]
	\item{Apodization mask:
        As described in \refsec{bb_maps}, the E-mode and $\phi$ maps that are used to form the B-mode template and the measured B-mode map are all apodized with the same mask.
	Since the coherence scale of lensing is $\sim 2$ degrees and the average deflection is $\sim 2$ arcmin, the B-mode map contains modes that are generated from E modes that lie outside the apodization mask.
	In this test, we extended the apodization mask used for $E_{\rm true}$ and $\phi_{\rm true}$ to cover at least 1 degree beyond the nominal B-mode mask edge.
	We then generate the B-mode template in the extended region then mask that template with the nominal mask before passing it to the delensing step.
	The delensing efficiency obtained from using the extended mask inputs is further improved to 98.5\%, from 95\%.
        For this reason these masking effects cannot be ignored for future measurements where the delensing residuals are required to be $<10\%$~\citep[e.g.][]{cmbs4scibk}.
	 }
	\item{Beyond $\mathcal{O}(\phi)$ lensing:
          The simulated lensing B-mode maps are generated from lensing Q/U maps with \textsc{Lenspix}, which includes higher-order lensing terms $\mathcal{O}(\phi^2)$.
	However, the quadratic estimator used to construct lensing B-mode templates with an unlensed E-mode map only includes terms up to order $\mathcal{O}(\phi)$.
	Part of the remaining residual power is due to higher-order $\mathcal{O}(\phi^2)$ effects in lensing that are not reconstructed in our estimator.
	\cite{2005PhRvD..71j3010C} compared the difference in the BB power spectrum between a full-sky perturbative calculations up to $\mathcal{O}(C_{\ell}^{\phi \phi})$ using a non-perturbative approach and found a difference of $\lesssim2\%$ for $\ell < 2500$.
	This effect contributes partially to the discrepancy between the delensing efficiency we obtain through our simulations and those from the analytic expression, where no lensing effects beyond $\mathcal{O}(\phi)$ are included.
	}

\end{enumerate}

Detailed studies are required in order to fully simulate delensing with realistic experimental non-idealities to the accuracy required by future experiments.
For example, iterative and maximum-likelihood methods for lensing B mode reconstruction may be needed to achieve the level of delensing required for future CMB experiments.

In addition, the expected constraints on inflation from future experiments will require CMB-internal delensing.
While biases are introduced by delensing with a CMB-reconstructed $\phi$, it has been shown that they can be sufficiently modeled and removed (\citealt{carron_internal_2017,namikawa_cmb_2017}).
These and other details are left to future work.

\section{Conclusion}
\label{sec:conclusion}

In this work, we have presented a demonstration of CMB B mode delensing.
We delens B-mode maps from \sptpol observations between 2012-2013 in the 95GHz and 150GHz bands.
We do so by subtracting a lensing B-mode template constructed from the \sptpol 150GHz CMB E-mode maps and the \textit{Herschel} $500\,\micron$ map of the cosmic infrared background (CIB) as the lensing potential tracer.

After subtracting our lensing B-mode template, the best-fit amplitude of the $C_{\ell}^{BB}$ power spectrum is reduced from \AlensNom\ to \AlensDel,
a \effData\% reduction in power.
The observed power reduction is found to be consistent with our realistic simulations, which, on average, remove 23\% of the lensing power through the delensing procedure.
We quantify the statistical significance of the reduction in power using the power spectrum difference, \Dclbb.
This level of removed power corresponds to a $\delSignificance\sigma$ significance rejection of the no-delensing hypothesis.
Furthermore, we find that this reduction in power is robust and not easily mimicked by instrumental systematics either in the CMB or the CIB data used in this work, 
\reftext{and it is robust against foreground modeling choices.}

Delensing will soon be essential for improving constraints on inflation from B-mode polarization signals.
Therefore we use our simulations to study how different factors affect the achieved delensing efficiency.
We find that the delensing efficiency of this analysis is limited primarily by noise in the estimated lensing potential from the CIB.
Fortunately, lensing potential estimates are expected to improve significantly in the near future;
Stage-3 CMB experiments will produce significantly lower-noise, multi-frequency CMB maps that will improve internal CMB lensing reconstructions.
We also find that at the level of delensing needed for CMB-S4, modes removed by timestream filtering can degrade delensing efficiency at a level that must be accounted for in instrument and survey-strategy planning.
Finally, we show that a naive application of the standard analytic delensing forecasting over-predicts the achieved delensing efficiency, highlighting the need for more realistic tools.

This demonstration of delensing the CMB B-mode spectrum is a first step towards the main purpose of this technique, namely improving constraints on primordial gravitational waves.
The 90 deg$^2$ of \sptpol data analyzed here would not place a meaningful constraint on the tensor-to-scalar ratio $r$, with or without delensing, particularly on the range of angular scales used in this work.
In the very near future, however, improved constraints on $r$ will come from delensing; in particular, work is ongoing to delens BICEP/Keck data with data from \sptpol and SPT-3G \citep{benson14}---possible because these experiments observe overlapping areas of sky.
On a longer timescale, CMB-S4 will produce significantly better CMB data, enabling and indeed requiring further improvements to the delensing procedure.

\acknowledgements{
The South Pole Telescope program is supported by the National Science Foundation through grant PLR-1248097.
Partial support is also provided by the NSF Physics Frontier Center grant PHY-0114422 to the Kavli Institute of Cosmological Physics at the University of Chicago, the Kavli Foundation, and the Gordon and Betty Moore Foundation through Grant GBMF\#947 to the University of Chicago.
Work at Argonne National Lab is supported by UChicago Argonne, LLC, Operator of Argonne National Laboratory (Argonne).
Argonne, a U.S. Department of Energy Office of Science Laboratory, is operated under Contract No. DE-AC02-06CH11357.
We also acknowledge support from the Argonne Center for Nanoscale Materials.
The McGill authors acknowledge funding from the Natural Sciences and Engineering Research Council of Canada, Canadian Institute for Advanced Research, and Canada Research Chairs program.
This work is also supported by the U.S. Department of Energy.
The CU Boulder group acknowledges support from NSF AST-0956135.
BB is supported by the Fermi Research Alliance, LLC under Contract No. De-AC02-07CH11359 with the U.S. Department of Energy.
CR acknowledges support from a Australian Research CouncilÕs Future Fellowship (FT150100074).
WLKW is grateful for support from the Croucher Foundation and hospitality of Stanford University.
JWH is supported by the National Science Foundation under Award No. AST-1402161.
The data analysis pipeline uses the scientific python stack \citep{hunter07, jones01, vanDerWalt11} and the HDF5 file format \citep{hdf5}.
The authors acknowledge useful discussions with Blake Sherwin and Toshiya Namikawa
and use of code first written by Duncan Hanson.
}

\bibliographystyle{apj}
\bibliography{../../BIBTEX/spt,./delens100d}

\begin{thebibliography}{}
\expandafter\ifx\csname natexlab\endcsname\relax\def\natexlab#1{#1}\fi

\bibitem[{{Abazajian} {et~al.}(2016){Abazajian}, {Adshead}, {Ahmed}, {Allen},
  {Alonso}, {Arnold}, {Baccigalupi}, {Bartlett}, {Battaglia}, {Benson},
  {Bischoff}, {Borrill}, {Buza}, {Calabrese}, {Caldwell}, {Carlstrom}, {Chang},
  {Crawford}, {Cyr-Racine}, {De Bernardis}, {de Haan}, {di Serego Alighieri},
  {Dunkley}, {Dvorkin}, {Errard}, {Fabbian}, {Feeney}, {Ferraro}, {Filippini},
  {Flauger}, {Fuller}, {Gluscevic}, {Green}, {Grin}, {Grohs}, {Henning},
  {Hill}, {Hlozek}, {Holder}, {Holzapfel}, {Hu}, {Huffenberger}, {Keskitalo},
  {Knox}, {Kosowsky}, {Kovac}, {Kovetz}, {Kuo}, {Kusaka}, {Le Jeune}, {Lee},
  {Lilley}, {Loverde}, {Madhavacheril}, {Mantz}, {Marsh}, {McMahon},
  {Meerburg}, {Meyers}, {Miller}, {Munoz}, {Nguyen}, {Niemack}, {Peloso},
  {Peloton}, {Pogosian}, {Pryke}, {Raveri}, {Reichardt}, {Rocha}, {Rotti},
  {Schaan}, {Schmittfull}, {Scott}, {Sehgal}, {Shandera}, {Sherwin}, {Smith},
  {Sorbo}, {Starkman}, {Story}, {van Engelen}, {Vieira}, {Watson}, {Whitehorn},
  \& {Kimmy Wu}}]{cmbs4scibk}
{Abazajian}, K.~N., {Adshead}, P., {Ahmed}, Z., {et~al.} 2016, ArXiv e-prints,
  arXiv:1610.02743

\bibitem[{Addison {et~al.}(2012)Addison, Dunkley, Hajian, Viero, Bond,
  {et~al.}}]{Addison:2011se}
Addison, G.~E., Dunkley, J., Hajian, A., {et~al.} 2012, Astrophys.J., 752, 120

\bibitem[{{Benson} {et~al.}(2014){Benson}, {Ade}, {Ahmed}, {Allen}, {Arnold},
  {Austermann}, {Bender}, {Bleem}, {Carlstrom}, {Chang}, {Cho}, {Cliche},
  {Crawford}, {Cukierman}, {de Haan}, {Dobbs}, {Dutcher}, {Everett}, {Gilbert},
  {Halverson}, {Hanson}, {Harrington}, {Hattori}, {Henning}, {Hilton},
  {Holder}, {Holzapfel}, {Irwin}, {Keisler}, {Knox}, {Kubik}, {Kuo}, {Lee},
  {Leitch}, {Li}, {McDonald}, {Meyer}, {Montgomery}, {Myers}, {Natoli},
  {Nguyen}, {Novosad}, {Padin}, {Pan}, {Pearson}, {Reichardt}, {Ruhl},
  {Saliwanchik}, {Simard}, {Smecher}, {Sayre}, {Shirokoff}, {Stark}, {Story},
  {Suzuki}, {Thompson}, {Tucker}, {Vanderlinde}, {Vieira}, {Vikhlinin}, {Wang},
  {Yefremenko}, \& {Yoon}}]{benson14}
{Benson}, B.~A., {Ade}, P.~A.~R., {Ahmed}, Z., {et~al.} 2014, in Society of
  Photo-Optical Instrumentation Engineers (SPIE) Conference Series, Vol. 9153,
  Society of Photo-Optical Instrumentation Engineers (SPIE) Conference Series,
  1

\bibitem[{B{\'e}thermin {et~al.}(2013)B{\'e}thermin, Wang, Dor{\'e}, Lagache,
  Sargent, {et~al.}}]{Bethermin:2013nza}
B{\'e}thermin, M., Wang, L., Dor{\'e}, O., {et~al.} 2013, arXiv:1304.3936

\bibitem[{{BICEP2 Collaboration}(2014)}]{bicep2a}
{BICEP2 Collaboration}. 2014, Physical Review Letters, 112, 241101

\bibitem[{{Carlstrom} {et~al.}(2011){Carlstrom}, {Ade}, {Aird}, {Benson},
  {Bleem}, {Busetti}, {Chang}, {Chauvin}, {Cho}, {Crawford}, {Crites}, {Dobbs},
  {Halverson}, {Heimsath}, {Holzapfel}, {Hrubes}, {Joy}, {Keisler}, {Lanting},
  {Lee}, {Leitch}, {Leong}, {Lu}, {Lueker}, {Luong-van}, {McMahon}, {Mehl},
  {Meyer}, {Mohr}, {Montroy}, {Padin}, {Plagge}, {Pryke}, {Ruhl}, {Schaffer},
  {Schwan}, {Shirokoff}, {Spieler}, {Staniszewski}, {Stark}, {Tucker},
  {Vanderlinde}, {Vieira}, \& {Williamson}}]{carlstrom11}
{Carlstrom}, J.~E., {Ade}, P.~A.~R., {Aird}, K.~A., {et~al.} 2011, \pasp, 123,
  568

\bibitem[{{Carron} {et~al.}(2017){Carron}, {Lewis}, \&
  {Challinor}}]{carron:2017}
{Carron}, J., {Lewis}, A., \& {Challinor}, A. 2017, \jcap, 5, 035

\bibitem[{Carron {et~al.}(2017)Carron, Lewis, \&
  Challinor}]{carron_internal_2017}
Carron, J., Lewis, A., \& Challinor, A. 2017, arXiv:1701.01712 [astro-ph],
  arXiv: 1701.01712

\bibitem[{{Challinor} \& {Lewis}(2005)}]{2005PhRvD..71j3010C}
{Challinor}, A., \& {Lewis}, A. 2005, \prd, 71, 103010

\bibitem[{{Crites} {et~al.}(2015){Crites}, {Henning}, {Ade}, {Aird},
  {Austermann}, {Beall}, {Bender}, {Benson}, {Bleem}, {Carlstrom}, {Chang},
  {Chiang}, {Cho}, {Citron}, {Crawford}, {de Haan}, {Dobbs}, {Everett},
  {Gallicchio}, {Gao}, {George}, {Gilbert}, {Halverson}, {Hanson},
  {Harrington}, {Hilton}, {Holder}, {Holzapfel}, {Hoover}, {Hou}, {Hrubes},
  {Huang}, {Hubmayr}, {Irwin}, {Keisler}, {Knox}, {Lee}, {Leitch}, {Li},
  {Liang}, {Luong-Van}, {McMahon}, {Mehl}, {Meyer}, {Mocanu}, {Montroy},
  {Natoli}, {Nibarger}, {Novosad}, {Padin}, {Pryke}, {Reichardt}, {Ruhl},
  {Saliwanchik}, {Sayre}, {Schaffer}, {Smecher}, {Stark}, {Story}, {Tucker},
  {Vanderlinde}, {Vieira}, {Wang}, {Whitehorn}, {Yefremenko}, \&
  {Zahn}}]{crites14}
{Crites}, A.~T., {Henning}, J.~W., {Ade}, P.~A.~R., {et~al.} 2015, \apj, 805,
  36

\bibitem[{{George} {et~al.}(2015){George}, {Reichardt}, {Aird}, {Benson},
  {Bleem}, {Carlstrom}, {Chang}, {Cho}, {Crawford}, {Crites}, {de Haan},
  {Dobbs}, {Dudley}, {Halverson}, {Harrington}, {Holder}, {Holzapfel}, {Hou},
  {Hrubes}, {Keisler}, {Knox}, {Lee}, {Leitch}, {Lueker}, {Luong-Van},
  {McMahon}, {Mehl}, {Meyer}, {Millea}, {Mocanu}, {Mohr}, {Montroy}, {Padin},
  {Plagge}, {Pryke}, {Ruhl}, {Schaffer}, {Shaw}, {Shirokoff}, {Spieler},
  {Staniszewski}, {Stark}, {Story}, {van Engelen}, {Vanderlinde}, {Vieira},
  {Williamson}, \& {Zahn}}]{george14}
{George}, E.~M., {Reichardt}, C.~L., {Aird}, K.~A., {et~al.} 2015, \apj, 799,
  177

\bibitem[{Griffin {et~al.}(2010)Griffin, Abergel, Abreu, Ade, Andr{\'e},
  {et~al.}}]{Griffin:2010hp}
Griffin, M., Abergel, A., Abreu, A., {et~al.} 2010, Astron.Astrophys., 518, L3

\bibitem[{{Hall} {et~al.}(2010){Hall}, {Keisler}, {Knox}, {Reichardt}, {Ade},
  {Aird}, {Benson}, {Bleem}, {Carlstrom}, {Chang}, {Cho}, {Crawford}, {Crites},
  {de Haan}, {Dobbs}, {George}, {Halverson}, {Holder}, {Holzapfel}, {Hrubes},
  {Joy}, {Lee}, {Leitch}, {Lueker}, {McMahon}, {Mehl}, {Meyer}, {Mohr},
  {Montroy}, {Padin}, {Plagge}, {Pryke}, {Ruhl}, {Schaffer}, {Shaw},
  {Shirokoff}, {Spieler}, {Stalder}, {Staniszewski}, {Stark}, {Switzer},
  {Vanderlinde}, {Vieira}, {Williamson}, \& {Zahn}}]{hall:2010}
{Hall}, N.~R., {Keisler}, R., {Knox}, L., {et~al.} 2010, \apj, 718, 632

\bibitem[{Hanson {et~al.}(2013)}]{hanson13}
Hanson, D., {et~al.} 2013, Phys.Rev.Lett., 111, 141301

\bibitem[{{Hivon} {et~al.}(2002){Hivon}, {G{\'o}rski}, {Netterfield}, {Crill},
  {Prunet}, \& {Hansen}}]{hivon02}
{Hivon}, E., {G{\'o}rski}, K.~M., {Netterfield}, C.~B., {et~al.} 2002, \apj,
  567, 2

\bibitem[{Holder {et~al.}(2013)Holder, Viero, Zahn, Aird, Benson,
  {et~al.}}]{Holder:2013hqu}
Holder, G., Viero, M., Zahn, O., {et~al.} 2013, Astrophys.J., 771, L16

\bibitem[{{Hu} \& {Dodelson}(2002)}]{hu02b}
{Hu}, W., \& {Dodelson}, S. 2002, \araa, 40, 171

\bibitem[{Hunter(2007)}]{hunter07}
Hunter, J.~D. 2007, Computing In Science \& Engineering, 9, 90

\bibitem[{Jones {et~al.}(2001)Jones, Oliphant, Peterson, {et~al.}}]{jones01}
Jones, E., Oliphant, T., Peterson, P., {et~al.} 2001, {SciPy}: Open source
  scientific tools for {Python}, [Online; accessed 2014-10-22]

\bibitem[{{Kamionkowski} \& {Kovetz}(2016)}]{kamionkowski15}
{Kamionkowski}, M., \& {Kovetz}, E.~D. 2016, \araa, 54, 227

\bibitem[{{Keck Array \& BICEP2 Collaborations} {et~al.}(2016){Keck Array \&
  BICEP2 Collaborations}, {Ade}, {Ahmed}, {Aikin}, {Alexander}, {Barkats},
  {Benton}, {Bischoff}, {Bock}, {Bowens-Rubin}, {Brevik}, {Buder}, {Bullock},
  {Buza}, {Connors}, {Crill}, {Duband}, {Dvorkin}, {Filippini}, {Fliescher},
  {Grayson}, {Halpern}, {Harrison}, {Hilton}, {Hui}, {Irwin}, {Karkare},
  {Karpel}, {Kaufman}, {Keating}, {Kefeli}, {Kernasovskiy}, {Kovac}, {Kuo},
  {Leitch}, {Lueker}, {Megerian}, {Netterfield}, {Nguyen}, {O'Brient},
  {Ogburn}, {Orlando}, {Pryke}, {Richter}, {Schwarz}, {Sheehy}, {Staniszewski},
  {Steinbach}, {Sudiwala}, {Teply}, {Thompson}, {Tolan}, {Tucker}, {Turner},
  {Vieregg}, {Weber}, {Wiebe}, {Willmert}, {Wong}, {Wu}, \& {Yoon}}]{bk14}
{Keck Array \& BICEP2 Collaborations}, {Ade}, P.~A.~R., {Ahmed}, Z., {et~al.}
  2016, Physical Review Letters, 116, 031302

\bibitem[{{Keisler} {et~al.}(2015){Keisler}, {Hoover}, {Harrington}, {Henning},
  {Ade}, {Aird}, {Austermann}, {Beall}, {Bender}, {Benson}, {Bleem},
  {Carlstrom}, {Chang}, {Chiang}, {Cho}, {Citron}, {Crawford}, {Crites}, {de
  Haan}, {Dobbs}, {Everett}, {Gallicchio}, {Gao}, {George}, {Gilbert},
  {Halverson}, {Hanson}, {Hilton}, {Holder}, {Holzapfel}, {Hou}, {Hrubes},
  {Huang}, {Hubmayr}, {Irwin}, {Knox}, {Lee}, {Leitch}, {Li}, {Luong-Van},
  {Marrone}, {McMahon}, {Mehl}, {Meyer}, {Mocanu}, {Natoli}, {Nibarger},
  {Novosad}, {Padin}, {Pryke}, {Reichardt}, {Ruhl}, {Saliwanchik}, {Sayre},
  {Schaffer}, {Shirokoff}, {Smecher}, {Stark}, {Story}, {Tucker},
  {Vanderlinde}, {Vieira}, {Wang}, {Whitehorn}, {Yefremenko}, \&
  {Zahn}}]{keisler15}
{Keisler}, R., {Hoover}, S., {Harrington}, N., {et~al.} 2015, \apj, 807, 151

\bibitem[{{Kesden} {et~al.}(2002){Kesden}, {Cooray}, \&
  {Kamionkowski}}]{kesden2002}
{Kesden}, M., {Cooray}, A., \& {Kamionkowski}, M. 2002, Physical Review
  Letters, 89, 011304

\bibitem[{Knox \& Song(2002)}]{knox2002}
Knox, L., \& Song, Y.-S. 2002, Phys. Rev. Lett., 89, 011303

\bibitem[{{Larsen} {et~al.}(2016){Larsen}, {Challinor}, {Sherwin}, \&
  {Mak}}]{larsen:2016}
{Larsen}, P., {Challinor}, A., {Sherwin}, B.~D., \& {Mak}, D. 2016, Physical
  Review Letters, 117, 151102

\bibitem[{{Lewis}(2005)}]{lewis05}
{Lewis}, A. 2005, \prd, 71, 083008

\bibitem[{{Lewis} \& {Bridle}(2002)}]{lewis02b}
{Lewis}, A., \& {Bridle}, S. 2002, \prd, 66, 103511

\bibitem[{{Mocanu} {et~al.}(2013){Mocanu}, {Crawford}, {Vieira}, {Aird},
  {Aravena}, {Austermann}, {Benson}, {B{\'e}thermin}, {Bleem}, {Bothwell},
  {Carlstrom}, {Chang}, {Chapman}, {Cho}, {Crites}, {de Haan}, {Dobbs},
  {Everett}, {George}, {Halverson}, {Harrington}, {Hezaveh}, {Holder},
  {Holzapfel}, {Hoover}, {Hrubes}, {Keisler}, {Knox}, {Lee}, {Leitch},
  {Lueker}, {Luong-Van}, {Marrone}, {McMahon}, {Mehl}, {Meyer}, {Mohr},
  {Montroy}, {Natoli}, {Padin}, {Plagge}, {Pryke}, {Rest}, {Reichardt}, {Ruhl},
  {Sayre}, {Schaffer}, {Shirokoff}, {Spieler}, {Spilker}, {Stalder},
  {Staniszewski}, {Stark}, {Story}, {Switzer}, {Vanderlinde}, \&
  {Williamson}}]{mocanu13}
{Mocanu}, L.~M., {Crawford}, T.~M., {Vieira}, J.~D., {et~al.} 2013, \apj, 779,
  61

\bibitem[{Namikawa(2017)}]{namikawa_cmb_2017}
Namikawa, T. 2017, arXiv:1703.00169 [astro-ph], arXiv: 1703.00169

\bibitem[{Padin {et~al.}(2008)Padin, Staniszewski, Keisler, Joy, Stark, Ade,
  Aird, Benson, Bleem, Carlstrom, Chang, Crawford, Crites, Dobbs, Halverson,
  Heimsath, Hills, Holzapfel, Lawrie, Lee, Leitch, Leong, Lu, Lueker, McMahon,
  Meyer, Mohr, Montroy, Plagge, Pryke, Ruhl, Schaffer, Shirokoff, Spieler, \&
  Vieira}]{padin08}
Padin, S., Staniszewski, Z., Keisler, R., {et~al.} 2008, Appl. Opt., 47, 4418

\bibitem[{Pilbratt {et~al.}(2010)Pilbratt, Riedinger, Passvogel, Crone, Doyle,
  {et~al.}}]{Pilbratt:2010mv}
Pilbratt, G., Riedinger, J., Passvogel, T., {et~al.} 2010, Astron.Astrophys.,
  518, L1

\bibitem[{{Planck Collaboration} {et~al.}(2014{\natexlab{a}}){Planck
  Collaboration}, {Ade}, {Aghanim}, {Armitage-Caplan}, {Arnaud}, {Ashdown},
  {Atrio-Barandela}, {Aumont}, {Baccigalupi}, {Banday}, \&
  et~al.}]{planck2013XVI}
{Planck Collaboration}, {Ade}, P.~A.~R., {Aghanim}, N., {et~al.}
  2014{\natexlab{a}}, \aap, 571, A16

\bibitem[{{Planck Collaboration} {et~al.}(2014{\natexlab{b}}){Planck
  Collaboration}, {Ade}, {Aghanim}, {Armitage-Caplan}, {Arnaud}, {Ashdown},
  {Atrio-Barandela}, {Aumont}, {Baccigalupi}, {Banday}, \&
  et~al.}]{planck2013XVIII}
---. 2014{\natexlab{b}}, \aap, 571, A18

\bibitem[{{Planck Collaboration} {et~al.}(2016{\natexlab{a}}){Planck
  Collaboration}, {Aghanim}, {Arnaud}, {Ashdown}, {Aumont}, {Baccigalupi},
  {Banday}, {Barreiro}, {Bartlett}, {Bartolo}, \& et~al.}]{planck15-11}
{Planck Collaboration}, {Aghanim}, N., {Arnaud}, M., {et~al.}
  2016{\natexlab{a}}, \aap, 594, A11

\bibitem[{{Planck Collaboration} {et~al.}(2016{\natexlab{b}}){Planck
  Collaboration}, {Ade}, {Aghanim}, {Arnaud}, {Ashdown}, {Aumont},
  {Baccigalupi}, {Banday}, {Barreiro}, {Bartlett}, \& et~al.}]{planck2015XV}
{Planck Collaboration}, {Ade}, P.~A.~R., {Aghanim}, N., {et~al.}
  2016{\natexlab{b}}, \aap, 594, A15

\bibitem[{{Planck Collaboration} {et~al.}(2016{\natexlab{c}}){Planck
  Collaboration}, {Adam}, {Ade}, {Aghanim}, {Arnaud}, {Aumont}, {Baccigalupi},
  {Banday}, {Barreiro}, {Bartlett}, \& et~al.}]{2016A&A...586A.133P}
{Planck Collaboration}, {Adam}, R., {Ade}, P.~A.~R., {et~al.}
  2016{\natexlab{c}}, \aap, 586, A133

\bibitem[{{POLARBEAR Collaboration}(2014{\natexlab{a}})}]{polarbear2014b}
{POLARBEAR Collaboration}. 2014{\natexlab{a}}, \apj, 794, 171

\bibitem[{{POLARBEAR Collaboration}(2014{\natexlab{b}})}]{polarbear2014c}
---. 2014{\natexlab{b}}, Physical Review Letters, 112, 131302

\bibitem[{{Polenta} {et~al.}(2005){Polenta}, {Marinucci}, {Balbi}, {de
  Bernardis}, {Hivon}, {Masi}, {Natoli}, \& {Vittorio}}]{polenta05}
{Polenta}, G., {Marinucci}, D., {Balbi}, A., {et~al.} 2005, Journal of
  Cosmology and Astro-Particle Physics, 11, 1

\bibitem[{{Seiffert} {et~al.}(2007){Seiffert}, {Borys}, {Scott}, \&
  {Halpern}}]{seiffert07}
{Seiffert}, M., {Borys}, C., {Scott}, D., \& {Halpern}, M. 2007, \mnras, 374,
  409

\bibitem[{{Seljak} \& {Hirata}(2004)}]{seljak2003}
{Seljak}, U., \& {Hirata}, C.~M. 2004, \prd, 69, 043005

\bibitem[{{Shaw} {et~al.}(2010){Shaw}, {Nagai}, {Bhattacharya}, \&
  {Lau}}]{shaw10}
{Shaw}, L.~D., {Nagai}, D., {Bhattacharya}, S., \& {Lau}, E.~T. 2010, \apj,
  725, 1452

\bibitem[{{Sherwin} \& {Schmittfull}(2015)}]{sherwin15}
{Sherwin}, B.~D., \& {Schmittfull}, M. 2015, \prd, 92, 043005

\bibitem[{{Simard} {et~al.}(2015){Simard}, {Hanson}, \& {Holder}}]{simard:2015}
{Simard}, G., {Hanson}, D., \& {Holder}, G. 2015, \apj, 807, 166

\bibitem[{{Smith} {et~al.}(2012){Smith}, {Hanson}, {LoVerde}, {Hirata}, \&
  {Zahn}}]{smith:2012}
{Smith}, K.~M., {Hanson}, D., {LoVerde}, M., {Hirata}, C.~M., \& {Zahn}, O.
  2012, \jcap, 6, 014

\bibitem[{{Smith} \& {Zaldarriaga}(2007)}]{smith07b}
{Smith}, K.~M., \& {Zaldarriaga}, M. 2007, \prd, 76, 043001

\bibitem[{{Smith} {et~al.}(2009){Smith}, {Cooray}, {Das}, {Dor{\'e}}, {Hanson},
  {Hirata}, {Kaplinghat}, {Keating}, {Loverde}, {Miller}, {Rocha}, {Shimon}, \&
  {Zahn}}]{smith08}
{Smith}, K.~M., {Cooray}, A., {Das}, S., {et~al.} 2009, in American Institute
  of Physics Conference Series, Vol. 1141, American Institute of Physics
  Conference Series, ed. S.~{Dodelson}, D.~{Baumann}, A.~{Cooray},
  J.~{Dunkley}, A.~{Fraisse}, M.~G. {Jackson}, A.~{Kogut}, L.~{Krauss},
  M.~{Zaldarriaga}, \& K.~{Smith}, 121--178

\bibitem[{Song {et~al.}(2003)Song, Cooray, Knox, \& Zaldarriaga}]{Song:2002sg}
Song, Y.-S., Cooray, A., Knox, L., \& Zaldarriaga, M. 2003, Astrophys.J., 590,
  664

\bibitem[{{Story} {et~al.}(2013){Story}, {Reichardt}, {Hou}, {Keisler}, {Aird},
  {Benson}, {Bleem}, {Carlstrom}, {Chang}, {Cho}, {Crawford}, {Crites}, {de
  Haan}, {Dobbs}, {Dudley}, {Follin}, {George}, {Halverson}, {Holder},
  {Holzapfel}, {Hoover}, {Hrubes}, {Joy}, {Knox}, {Lee}, {Leitch}, {Lueker},
  {Luong-Van}, {McMahon}, {Mehl}, {Meyer}, {Millea}, {Mohr}, {Montroy},
  {Padin}, {Plagge}, {Pryke}, {Ruhl}, {Sayre}, {Schaffer}, {Shaw}, {Shirokoff},
  {Spieler}, {Staniszewski}, {Stark}, {van Engelen}, {Vanderlinde}, {Vieira},
  {Williamson}, \& {Zahn}}]{story13}
{Story}, K.~T., {Reichardt}, C.~L., {Hou}, Z., {et~al.} 2013, \apj, 779, 86

\bibitem[{{Story} {et~al.}(2015){Story}, {Hanson}, {Ade}, {Aird}, {Austermann},
  {Beall}, {Bender}, {Benson}, {Bleem}, {Carlstrom}, {Chang}, {Chiang}, {Cho},
  {Citron}, {Crawford}, {Crites}, {de Haan}, {Dobbs}, {Everett}, {Gallicchio},
  {Gao}, {George}, {Gilbert}, {Halverson}, {Harrington}, {Henning}, {Hilton},
  {Holder}, {Holzapfel}, {Hoover}, {Hou}, {Hrubes}, {Huang}, {Hubmayr},
  {Irwin}, {Keisler}, {Knox}, {Lee}, {Leitch}, {Li}, {Liang}, {Luong-Van},
  {McMahon}, {Mehl}, {Meyer}, {Mocanu}, {Montroy}, {Natoli}, {Nibarger},
  {Novosad}, {Padin}, {Pryke}, {Reichardt}, {Ruhl}, {Saliwanchik}, {Sayre},
  {Schaffer}, {Smecher}, {Stark}, {Tucker}, {Vanderlinde}, {Vieira}, {Wang},
  {Whitehorn}, {Yefremenko}, \& {Zahn}}]{story14}
{Story}, K.~T., {Hanson}, D., {Ade}, P.~A.~R., {et~al.} 2015, \apj, 810, 50

\bibitem[{{The HDF Group}(1997)}]{hdf5}
{The HDF Group}. 1997, {Hierarchical Data Format, version 5},
  http://www.hdfgroup.org/HDF5/

\bibitem[{{Tristram} {et~al.}(2005){Tristram}, {Mac{\'{\i}}as-P{\'e}rez},
  {Renault}, \& {Santos}}]{tristram05}
{Tristram}, M., {Mac{\'{\i}}as-P{\'e}rez}, J.~F., {Renault}, C., \& {Santos},
  D. 2005, \mnras, 358, 833

\bibitem[{van~der Walt {et~al.}(2011)van~der Walt, Colbert, \&
  Varoquaux}]{vanDerWalt11}
van~der Walt, S., Colbert, S., \& Varoquaux, G. 2011, Computing in Science
  Engineering, 13, 22

\bibitem[{{van Engelen} {et~al.}(2015){van Engelen}, {Sherwin}, {Sehgal},
  {Addison}, {Allison}, {Battaglia}, {de Bernardis}, {Bond}, {Calabrese},
  {Coughlin}, {Crichton}, {Datta}, {Devlin}, {Dunkley}, {D{\"u}nner},
  {Gallardo}, {Grace}, {Gralla}, {Hajian}, {Hasselfield}, {Henderson}, {Hill},
  {Hilton}, {Hincks}, {Hlozek}, {Huffenberger}, {Hughes}, {Koopman},
  {Kosowsky}, {Louis}, {Lungu}, {Madhavacheril}, {Maurin}, {McMahon},
  {Moodley}, {Munson}, {Naess}, {Nati}, {Newburgh}, {Niemack}, {Nolta}, {Page},
  {Pappas}, {Partridge}, {Schmitt}, {Sievers}, {Simon}, {Spergel}, {Staggs},
  {Switzer}, {Ward}, \& {Wollack}}]{van-engelen:2015}
{van Engelen}, A., {Sherwin}, B.~D., {Sehgal}, N., {et~al.} 2015, \apj, 808, 7

\bibitem[{{Zaldarriaga} \& {Seljak}(1998)}]{zaldarriaga98}
{Zaldarriaga}, M., \& {Seljak}, U. 1998, \prd, 58, 23003 (6 pages)

\end{thebibliography}

\end{document}